\documentclass[11pt,a4paper,english]{article}

\usepackage{graphicx}
\usepackage{amssymb,latexsym}
\usepackage{amsmath}
\usepackage{epsf}
\usepackage{pdfsync}
\usepackage{epsfig}
\usepackage[parfill]{parskip}
\usepackage[matrix,arrow,color]{xy}
\usepackage[usenames]{color}
\usepackage{hyperref}
\usepackage{bbm}
\usepackage{dsfont}
\usepackage{stmaryrd}
\usepackage{xfrac}
\usepackage{cite}

\usepackage{marginnote}

\usepackage{upgreek}

\input{xy}
\xyoption{all}

\input{epsf}

\makeatletter

\catcode`@=11
\renewcommand{\section}
{\@startsection{section}{1}{0pt}{\medskipamount}{\medskipamount}{\large\bf}}
\makeatletter\renewcommand{\subsection}
{\@startsection{subsection}{2}{\z@}{-3.25ex plus -1ex minus -.2ex}
{1.5ex plus .2ex}{\it }}

\numberwithin{equation}{section}
\catcode`@=12

\newcommand{\ba}{\begin{eqnarray*}}
\newcommand{\ea}{\end{eqnarray*}}
\newcommand{\ban}{\begin{eqnarray}}
\newcommand{\ean}{\end{eqnarray}}

\newcommand{\cW}{{\cal W}}
\newcommand{\cN}{{\cal N}}
\newcommand{\cM}{{\cal M}}
\newcommand{\cS}{{\cal S}}

\newcommand{\cO}{{\cal O}}

\newcommand{\cR}{{\cal R}}

\newcommand{\cL}{{\cal L}}

\newcommand{\cP}{{\cal P}}

\newcommand{\mbf}[1]{{\boldsymbol {#1} }}
\newcommand{\real}{{\mathbb R}} 

\def\e{{\,\rm e}\,}
\newcommand{\wt}{\widetilde}

\def\ii{{\,{\rm i}\,}}
\def\dd{{\rm d}}

\def\beq{\begin{equation}}
\def\bee{\begin{equation}}
\def\eeq{\end{equation}}
\def\bea{\begin{eqnarray}}
\def\eea{\end{eqnarray}}
\def\bd{\begin{displaymath}}
\def\ed{\end{displaymath}}

\newcommand{\Cint}{\int\kern-10.5pt-\kern7pt}

\newcommand{\be}{\begin{equation}}
\newcommand{\ee}{\end{equation}}

\newcommand\fverbit{\egroup\item[\fbox{\unhbox\pippobox}]}
\newbox\pippobox

{

\def\bs{\bar{\sigma}}

\def\w{\wedge}

\def\rd{\overset{\shortrightarrow}{\partial}}
\def\ldelta{\overset{\shortleftarrow}{\delta}}
\def\rdelta{\overset{\shortrightarrow}{\delta}}

\def\tx{\widetilde{X}}

\def\bs#1{\boldsymbol{#1}}
\def\rm#1{\mathrm{#1}}
\def\bv#1#2{\left( {#1},{#2} \right)_{\mathrm{BV}} }
\def\wt#1{\widetilde{#1}}

\def\be{\begin{equation}}
\def\ee{\end{equation}}
\def\bea{\begin{eqnarray}}
\def\eea{\end{eqnarray}}

\begin{document}

\begin{titlepage}
\setcounter{page}{0}
\begin{flushright}
\small
{\sf EMPG--18--11}
\end{flushright}
\normalsize

\vskip 1.8cm

\begin{center}

{\Large\bf Double field theory for the A/B-models \\[2mm] and topological S-duality in generalized geometry}

\vspace{15mm}

{\large\bf Zolt\'an K\"ok\'enyesi$^{(a),(b)}$, Annam{\'a}ria Sinkovics$^{(b)}$ and
 Richard~J.~Szabo$^{(c)}$}
\\[6mm]
\noindent{\em $^{(a)}$ MTA Lend{\"u}let Holographic QFT Group\\ Wigner Research Center for
Physics\\
Konkoly-Thege Mikl\'os {\'u}t 29-33, 1121 Budapest, Hungary} \\
Email: \ {\tt
  kokenyesi.zoltan@wigner.mta.hu} \\[4mm]
\noindent{\em $^{(b)}$ Institute of Theoretical Physics\\ MTA-ELTE
  Theoretical Research Group \\ E\"otv\"os Lor\'and University \\ P\'azm\'any s. 1/A, 1117
  Budapest, Hungary} \\ Email: \ {\tt
  sinkovics@general.elte.hu}\\[4mm]
\noindent{\em $^{(c)}$ Department of Mathematics\\ Heriot-Watt
  University\\
Colin Maclaurin Building, Riccarton, Edinburgh EH14 4AS, UK\\ 
Maxwell Institute for Mathematical Sciences, Edinburgh, UK\\
The Higgs Centre for Theoretical Physics, Edinburgh, UK}\\
Email: \ {\tt
  R.J.Szabo@hw.ac.uk}

\vspace{20mm}

\begin{abstract}
\noindent
We study AKSZ-type BV constructions for the topological A- and
B-models within a double field theory formulation that incorporates
backgrounds with geometric and non-geometric fluxes. We relate them to a
Courant sigma-model, on an open membrane, corresponding to a
generalized complex structure, which reduces to the A- or B-models on
the boundary. We introduce S-duality at the level of the membrane
sigma-model based on the generalized complex structure, which
exchanges the related AKSZ field theories, and interpret it as topological S-duality of the A- and B-models. Our approach leads to new classes of Courant algebroids associated to (generalized) complex geometry.
\end{abstract}

\end{center}


\end{titlepage}



{\baselineskip=18pt
\tableofcontents
}

\newpage

\section{Introduction}

The topological A- and B-models were introduced originally by Witten~\cite{Witten1988a,Witten1988b,Witten1991} as twists of two-dimensional $\cN = 2$ supersymmetric sigma-models, and were subsequently used to construct topological string theory~\cite{Bershadsky1994} by coupling them to two-dimensional topological gravity through integration over the moduli space of the target Calabi-Yau manifold. 
In this paper we reformulate the AKSZ constructions of the A- and B-models in the framework of double field theory as a single membrane sigma-model, and study its relation to generalized complex geometry and S-duality.

AKSZ formulations provide natural geometric methods for constructing BV quantized sigma-models~\cite{AKSZ1997,Cattaneo2001,Bouwknegt:2011vn,Ikeda2012}, and they produce examples of topological field theories of Schwarz-type in arbitrary dimensionality such as the Poisson sigma-model, Chern-Simons theory and BF-theory; special gauge fixing action functionals also yield examples of topological field theories of Witten-type, such as the A- and B-models. 
One motivation for studying them comes from flux compactifications of
type~II string theory, where fluxes appear as twists of the Courant
algebroid on the T-duality inspired generalized tangent
bundle~\cite{Hull2005,Shelton2005,Blumenhagen2012,Heller2016}. Courant
algebroids are in one-to-one correspondence to three-dimensional AKSZ
sigma-models with target QP-manifold of degree 2, called Courant
sigma-models~\cite{Roytenberg2002,Park2000,Ikeda2003,Roytenberg2002b,Hofman2002,Hofman2002a,Roytenberg2007},
which geometrize fluxes in the sense that they are uplifts of string
sigma-models to one higher dimension that can accomodate the
fluxes~\cite{Halmagyi,Bouwknegt:2011vn,Ikeda2012,Richard2012,Chatzistavrakidis2015}
(see~\cite{Szabo2018} for a review). A
more natural framework for describing fluxes is double field theory
(DFT)~\cite{doubled3,Siegel:1993th,dft1}, where the original coordinates conjugate to closed string
momentum modes are extended with dual coordinates conjugate to winding
modes, and the continuous version of T-duality becomes a manifest
symmetry of the theory (see~\cite{Aldazabal2013,Berman2014,Hohm2013}
for reviews). The algebroid structure of double field theory was
studied
in~\cite{Hull:2009zb,Vaisman:2012ke,Deser2015,Deser2016,Heller2016,Freidel:2017yuv,Richard2018,Svoboda:2018rci};
in particular, the notion of DFT algebroid was introduced
in~\cite{Richard2018} whose derived bracket is the C-bracket of double
field theory, and whose corresponding membrane sigma-model naturally
captures the T-duality orbit of geometric and non-geometric flux
backgrounds in a single unified description.

The topological A- and B-model
string theories
in backgrounds with $H$-flux are captured by
generalized complex geometry~\cite{Kapustin:2003sg,Kapustin:2004gv,Pestun:2005rp}. 
They have been extensively studied by
introducing AKSZ string and open membrane sigma-models with
generalized complex structures, which reduce upon gauge fixing to the A- and B-models~\cite{Zucchini2004,Zucchini2005,Pestun2006,Ikeda2007,Stojevic2005}. The novelty of our approach is that their AKSZ sigma-models are reformulated within double field theory on both the string and membrane levels, which gives a more natural explanation of how their AKSZ constructions are related to generalized complex structures. It also highlights some new aspects, such as how topological S-duality appears on the level of AKSZ sigma-models and can be traced back to generalized complex geometry.

Based on our observation that the Poisson sigma-model on doubled
spaces captures both the A- and B-models with different choices of the
doubled Poisson structure, we propose an open AKSZ membrane
sigma-model, inspired by the approach of~\cite{Bessho2015} to T-duality
between geometric and non-geometric fluxes, which gives back the doubled Poisson sigma-model on the boundary in a specific gauge. Then we reduce the fields in a way which can be interpreted as the same reduction performed in~\cite{Richard2018}, where it was called a DFT projection. The resulting AKSZ membrane sigma-model captures a particular class of generalized complex structures given by an initial Poisson and complex structure. It therefore corresponds to a Courant algebroid for the generalized complex structure with the identities of its integrability condition; to the best of our knowledge this is a new example of Courant algebroid. We also show that the AKSZ membrane theory can be reduced through gauge fixing to the A- or B-models on the boundary if the generalized complex structure is given by a purely Poisson or complex structure respectively. Furthermore, we find a realization of topological S-duality, which exchanges the weakly and strongly coupled sectors of the topological A- and B-model string theories~\cite{Nekrasov2004b}, on the level of the AKSZ construction. Our result is based on an S-duality which maps Poisson and complex structure Courant algebroids into each other, and lies within the Courant algebroid for the generalized complex structure. This duality is promoted to the AKSZ membrane sigma-model and interpreted as S-duality which relates the couplings $g_{\rm A}$ and $g_{\rm B}$ of the A- and B-models in the usual way: $g_{\rm{A}}=1/g_{\rm{B}}$. 

\begin{figure}[tb]
	\centering
		\includegraphics[width=1.0\textwidth]{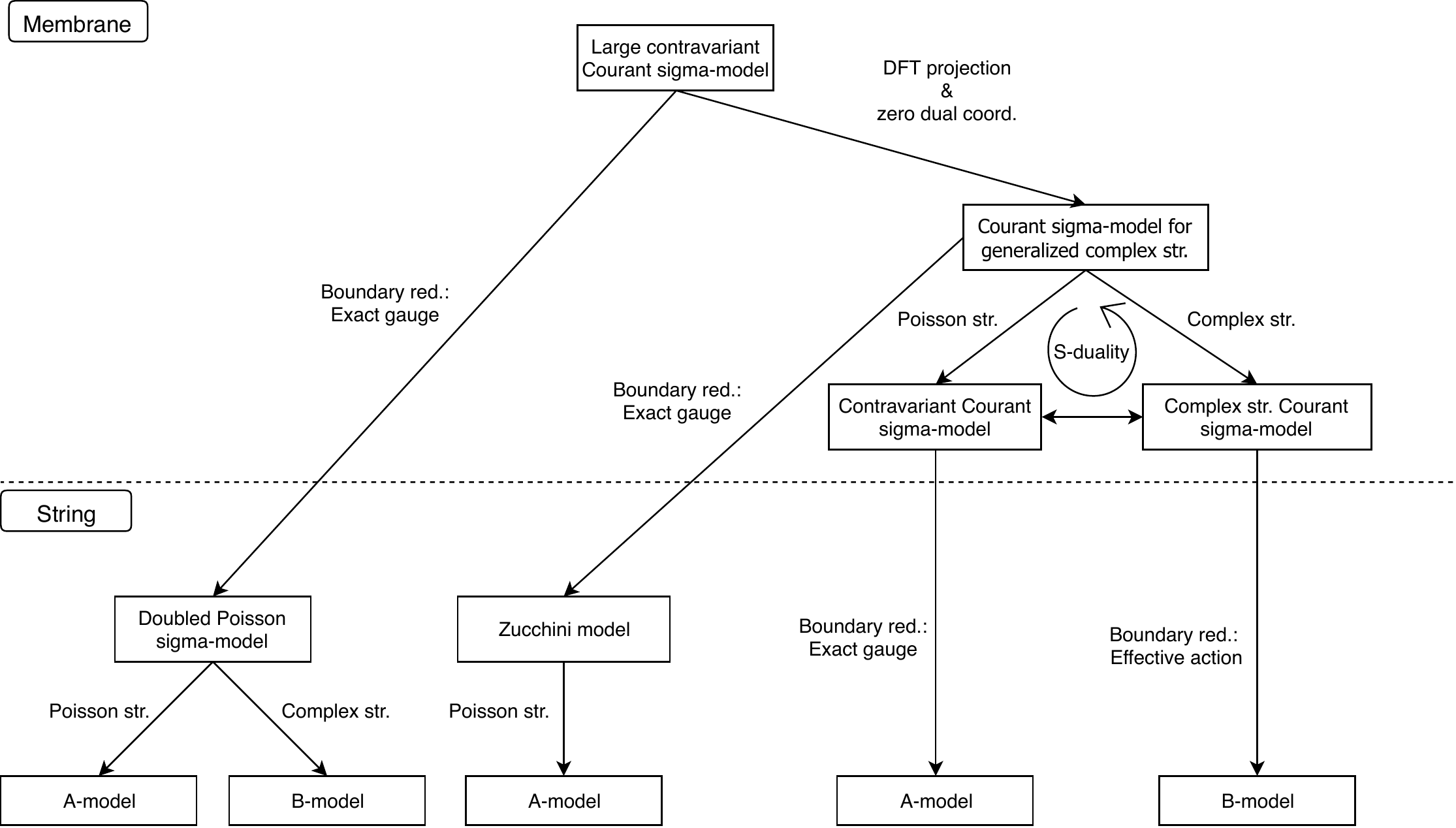}
	\caption{\small Schematic presentation of the different reductions and connections between the AKSZ string and membrane sigma-models related to the topological A- and B-models.}
	\label{fig:Diagram}
\end{figure}

In Figure~\ref{fig:Diagram} we summarize the relations between the
different AKSZ string and membrane sigma-models appearing in this
paper. In the following we also discuss some additional new results:
We introduce a gauge fixing method which is a natural way to
perform boundary reductions, and we show that the contravariant
Courant sigma-model of~\cite{Bessho2015} reduces to the Poisson
sigma-model in this gauge; we also relate its twisting by (geometric)
$R$-flux to the membrane sigma-model introduced in~\cite{Richard2012}
which describes the nonassociative deformations of (locally
non-geometric) $R$-flux backgrounds. We further show that the standard
Courant sigma-model~\cite{Ikeda2012} twisted by $H$-flux is related to the B-model in a particular gauge.

This paper is organized as follows. In~\S\ref{sec:AKSZBackground} we
give a relatively detailed overview of the AKSZ construction, together
with two techniques for dimensional reduction: the first method is new
and is specific to boundary reductions, whereas the second method is
based on the more general reductions through effective actions
discussed in~\cite{Cattaneo2009,Ikeda2007}. In~\S\ref{sec:StringAKSZ}
we survey some aspects of string sigma-models constructed by the AKSZ
method, and in particular the various AKSZ formulations of the A- and
B-models. In~\S\ref{sec:CourantSigmaMod} we review Courant
sigma-models together with two relevant examples, the standard and the
contravariant Courant sigma-model, and we describe their boundary
reductions. In~\S\ref{sec:DFTAlgAndSigmaModel} we recall the DFT
projection which we use in later sections to derive the Courant
sigma-model for generalized complex
geometry. In~\S\ref{sec:UnifyABDFT} we describe how to reformulate
the A- and B-models within double field theory, and
apply the DFT projection to show
how it reduces to the A- and B-models on the
boundary. In~\S\ref{sec:Sduality} we introduce the duality between the
Poisson and complex structure Courant algebroids, and connect it to
topological S-duality. In~\S\ref{sec:Conc} we close with some
concluding remarks, and directions for further study of our
geometrical structures and constructions, while in
Appendix~\ref{sec:appGradedFormulas} we summarize some key formulas
from the differential calculus of graded functionals relevant for the AKSZ construction which we have not found in the literature.

\section{AKSZ constructions and dimensional reductions}
\label{sec:AKSZBackground}

The AKSZ construction is a BV quantized sigma-model formulation which
produces a geometric solution to the classical master equation, called
the AKSZ action. In this section we survey some pertinent background about the AKSZ
construction and BV quantization; a more complete review can be found in~\cite{Ikeda2012}.
 We then introduce a new boundary reduction method using a BV gauge
 fixing procedure, and review another known dimensional reduction
 technique based on effective actions by integration over one
 coordinate direction which is not specific to boundary reductions.

\subsection{AKSZ sigma-models}
\label{sec:AKSZingeneral}

The basic ingredients of AKSZ theory consists of two classes of
supermanifolds. The `source' $(\cW,Q_\cW,\mu)$ consists of a
differential graded (dg-)manifold, which is a graded manifold $\cW$
equiped with a cohomological vector field $Q_\cW$, i.e. $Q_\cW$ is of degree~1 and its Lie derivative
$\cL_{Q_\cW}$ squares to zero, and a measure $\mu$ which is invariant under
$Q_\cW$. In this paper we take
$\cW=T[1]\Sigma_{d}$, the tangent bundle of a $d$-dimensional oriented worldvolume manifold $\Sigma_{d}$ with the degree of its
fibers shifted by~1, which is isomorphic to the exterior algebra of
differential forms $(\Omega(\Sigma_{d}),\wedge)$. We choose the
cohomological vector field $Q_\cW$ corresponding to the de Rham
differential, which in local affine coordinates $\hat
z^{\hat\mu}=(\sigma^\mu,\theta^\mu)\in T[1]\Sigma_{d}$, with degree~0
coordinates $\sigma^\mu$ on $\Sigma_d$ and degree~1 fiber coordinates
$\theta^\mu$, has the form $Q_\cW=\theta^\mu\,
\frac\partial{\partial\sigma^\mu}=:\mbf D$, where repeated upper and lower indices are always implicitly understood to be summed over.
 The measure in local coordinates can be written in the form
 $\mu=\dd^d\hat z:= \dd^{d} \sigma \ \dd^{d} \theta$. 

The `target' $(\cM,Q_\gamma,\omega)$ is a symplectic
dg-manifold, which is a graded manifold $\cM$ with a cohomological vector
field $Q_\gamma$, and a graded symplectic form $\omega$ for which
$Q_\gamma$ is a Hamiltonian vector field: $\iota_{Q_\gamma}\omega=\dd
\gamma$ for some Hamiltonian function $\gamma$ on $\cM$, where
$\iota_Q$ denotes contraction of a differential form along the vector
field $Q$. In order to reproduce the BV formalism, the symplectic structure $\omega$ is taken to be of degree $d+1$, so that the Hamiltonian function $\gamma$ is of degree $d$. 
A common choice of target for the AKSZ construction is to take $\cM$
to be an N-manifold, which is a graded manifold with no coordinates of
negative degree. In this case the triple $(\cM,Q_\gamma,\omega)$ is
called a QP-manifold of degree $n=d-1$, or QP$n$-manifold for
short. They typically arise from $n$-graded vector bundles over the
degree~0 body $M=\cM_0$ of $\cM$~\cite{Ikeda2012}, and in particular functions of degree $n-1$ can be identified with sections of a vector bundle $E\to M$ equiped with the structure of a Leibniz algebroid.
In subsequent sections we describe the AKSZ topological field theories
associated with the first two non-trivial members in the hierarchy of
QP-structures on the target manifold for dimensions $d=2,3$, in the
context of the string and membrane sigma-models of interest in this
paper. 
 
The AKSZ space of fields is the mapping space
\be
\mbf\cM = {\sf Map}\big(T[1]\Sigma_d\,,\,\cM\big)
\ee
consisting of smooth maps from $(T[1]\Sigma_d,\mbf D,\mu)$ to
$(\cM,Q_\gamma,\omega)$, which we refer to as superfields in the following. We can introduce local coordinates on $\mbf\cM$ via the superfields
\be
\hat{\mbf X}{}^{\hat\imath}(\hat z^{\hat\mu}) \, = \, \mbf\phi^*(\hat
X^{\hat\imath}\,)(\hat z^{\hat\mu}) \  ,
\ee
for local coordinates $\hat z^{\hat\mu}\in \cW$, $\hat
X^{\hat\imath}\in \cM$ and $\mbf\phi\in\mbf\cM$. 
The cohomological vector fields $Q_\cW=\mbf D$ and $Q_\gamma$ induce a
cohomological vector field $\bs{Q}$ on $\mbf\cM$ in the following
way. For $\mbf\phi\in \mbf\cM$ and $\hat z\in \cW$, use local
coordinates to define 
\be 
\bs{Q}_0  =  \int_{T[1]\Sigma_{d}}\, \dd^d \hat z \ \bs{D} \hat{\mbf X}{}^{\hat\imath}(\hat{z})\, \frac{\mbf\delta}{\mbf\delta\hat{\mbf X}{}^{\hat\imath}(\hat{z})}  \qquad \mbox{and} \qquad
\bs{Q}_\gamma =  \int_{T[1]\Sigma_{d}}\, \dd^d \hat z \ Q_\gamma^{\hat\imath} \big(\hat{\mbf X}{}(\hat{z})\big)\, \frac{\mbf\delta}{\mbf\delta\hat{\mbf X}{}^{\hat\imath}(\hat{z})} \ ,
\ee
where $Q_\gamma^{\hat\imath}(\hat{X}) \, \partial/ \partial
\hat X^{\hat\imath}$ is the local form of the vector field $Q_\gamma$
on $\cM$; relevant definitions and formulas in differential calculus on mapping superspaces are summarized in Appendix~\ref{sec:appGradedFormulas}.
Then $\mbf\cM$ is a dg-manifold with the cohomological vector field
\be
\bs{Q}  \, = \,  \bs{Q}_0 + \bs{Q}_\gamma \ . 
\ee
We note that $\bs{Q}_0$ has ghost number\footnote{We use the
  terminology `ghost number' for the degree of a superfield $\mbf\phi$ in $\mbf\cM$.} $|\bs{D}|-d=1-d$ and $\bs{Q}_\gamma$ has ghost number $|Q_\gamma|-d=1-d$ as well, where $|Q_\gamma|$ denotes the degree of $Q_\gamma$. If a vector field on $\mbf\cM$ acts as a derivative, its ghost number is shifted by $d-1$, because a vector field based with coordinate $\hat{\bs{X}}{}^{\hat\imath}(\hat{z})$ has ghost number $|\hat{X}{}^{\hat\imath}|-1$, but a functional derivative with respect to $\hat{\bs{X}}{}^{\hat\imath}(\hat{z})$ has ghost number $|\hat{X}{}^{\hat\imath}|+d$.

Given an $n$-form $\alpha\in\Omega^n(\cM)$, we can lift it to an $n$-form $\bs{\alpha}\in\Omega^n(\mbf\cM)$ by transgression to the mapping space as
\be
\bs{\alpha} \, = \, \int_{T[1]\Sigma_d}\, \dd^d \hat z \ \mathrm{ev}^* (\alpha) \ ,
\ee
where $\mathrm{ev}\,:\, T[1]\Sigma_d \times \mbf\cM \, \rightarrow \cM $ is the
evaluation map. As we see, $\bs{\alpha}$ is
an $n$-form functional of the fields in $\mbf\cM$,
and due to the integration $\bs{\alpha}$ has ghost number $|\alpha|-d$, where $|\alpha|$
denotes the total degree of $\alpha$ (i.e.~the form degree coming from
the grading of $\mbf\delta$ plus the degree of the graded coordinates). In particular, since
transgression is a chain map, from the degree $d+1$ symplectic form $\omega$ on $\cM$
and a Liouville potential $\vartheta$, such that
$\omega=\dd\vartheta$, we get the symplectic form $\bs{\omega}$ of
ghost number $1$ and Liouville potential $\bs{\vartheta}$ on $\mbf\cM$, such
that $\mbf\omega=\mbf\delta\mbf\vartheta$. Furthermore, the
cohomological vector field $\bs{Q}$ on $\mbf\cM$ is also Hamiltonian with
Hamiltonian function $-\iota_{\mbf Q_0}\mbf\vartheta + \bs{\gamma}$ of degree~0:
$\iota_{\bs{Q}_\gamma}\bs{\omega}= \bs{\delta} \bs{\gamma}$, where
$\iota_{\bs{Q}_0}$ and $\iota_{\bs{Q}_\gamma}$ have ghost number 0,
while $\bs{\gamma}$ has ghost number $|\gamma|-d=0$. In other words,
the mapping space of superfields $\mbf\cM$ is itself a symplectic dg-manifold.

The BV bracket $(\,\cdot\,,\,\cdot\,)_{\mathrm{BV}}$ is the graded Poisson bracket of ghost number~1 on $\mbf\cM$ defined from $\mbf\omega$, and it corresponds to the graded Poisson bracket $\{\,\cdot\,,\,\cdot\,\}$ of degree $-d+1$ on $\cM$ defined from $\omega$, since the transgression map $\int_{T[1]\Sigma_d}\, \dd^d \hat z \ \mathrm{ev}^*$ is a Lie algebra homomorphism from $(\cM,\{\,\cdot\,,\,\cdot\,\})$ to $(\mbf\cM,\bv{\,\cdot\,}{\,\cdot\,})$.
The cohomological vector fields as derivatives can be represented through derived
brackets as
\be
Q_\gamma=\{\gamma,\,\cdot\,\} \qquad \mbox{and} \qquad  \overset{\shortrightarrow}{\mbf Q}= \bv{\mbf \cS}{\,\cdot\,} \ ,
\ee
where the Hamiltonian $\mbf \cS$ on $\mbf\cM$ is defined to be the
AKSZ action, which is the desired BV action. To explicitly specify it,
we choose a Liouville potential $\vartheta$ on $\cM$, and consider its
zero locus $\cL$ which is a Lagrangian submanifold of $\cM$. We pick a
submanifold $\cL'\subseteq \cL $ and restrict the space of fields
$\mbf\cM$ to the subspace $\mbf\cM_{\cL'}\subset\mbf\cM$ consisting of
maps that send the boundary $\partial \cW = T[1]\partial \Sigma_{d}$
into $\cL'$. This assigns boundary conditions on our fields, and now
we can write the degree~0 AKSZ action $\mbf \cS$ on $\mbf\cM_{\cL'}$ in the form
\be\label{eq:AKSZactiongeneral}
\bs{\cS} \, = \, \bs{\cS}_0 \, + \, \bs{\gamma} \ ,
\ee 
where
\be
\bs{\cS}_0=-\iota_{\bs{Q}_0}\bs\vartheta 
\ee 
is the kinetic term and the Hamiltonian function 
\be
\mbf\gamma =  \int_{T[1]\Sigma_{d}}\,\dd^d\hat z \ \mathrm{ev}^*(\gamma)
\ee
is the interaction
term. 
The cocycle conditions $Q_\gamma^2=0$ and $ \overset{\shortrightarrow}{\mbf Q}{}^2=0$ are equivalent to $\{\gamma,\gamma\}=0$ and $\bv{\mbf\gamma}{\mbf\gamma}=0$, hence the AKSZ action is a solution of the classical master equation $\bv{\mbf\cS}{\mbf\cS}=0$.
In the BV formalism, the cohomological vector field $\overset{\shortrightarrow}{\mbf Q}$
corresponds to the BRST charge.

A canonical transformation is associated to a degree $d-1$ function
$\alpha$ on $\cM$. We use the notation $\delta_\alpha$ for the
corresponding Hamiltonian vector field, and 
$\e^{\delta_\alpha}$ and $\e^{\delta_{\bs{\alpha}}}$ for the
respective canonical transformations. The action of the canonical
transformation on $\bs{\gamma}$ is given by
$\e^{\delta_{\bs{\alpha}}}\bs{\gamma}=\int_{T[1]\Sigma_d}\,\dd^d\hat z \ \mathrm{ev}^*
(\e^{\delta_\alpha} \gamma)$, which preserves the classical master
equation as
\be
\{\e^{\delta_\alpha}\gamma,\e^{\delta_\alpha}\gamma\}
=\e^{\delta_\alpha}\{\gamma,\gamma\} = 0 \ ,
\ee
due to $\{\gamma,\gamma\}=0$. If $\alpha|_{\cL'}=0$, then the AKSZ
action $\bs{\cS}_0 \, + \, \bs{\gamma}$ is equivalent to $\bs{\cS}_0
\, + \, \e^{\delta_{\bs{\alpha}}}\bs{\gamma}$ up to a canonical
transformation. The canonical transformation
$\e^{\delta_{\bs{\alpha}}}$ is an example of a duality transformation,
which in the AKSZ formalism is defined to be a symplectomorphism,
i.e. a diffeomorphism between underlying symplectic manifolds
which preserves the symplectic structures.

We can then introduce a boundary term in the AKSZ action using the ingredients of a canonical transformation. Let $\beta$ be a degree $d-1$ function on $\cM$ as before, and further assume that $\{ \beta , \beta \} = 0$ and $\e^{\delta_\beta} \gamma\big|_{\cL'}=0$. Then the AKSZ action $\bs{\cS}_0 \, + \, \bs{\gamma}$ on $\mbf\cM_{\cL'}$ is equivalent to the AKSZ action 
\be
\bs{\cS} \, = \, \bs{\cS}_0 \, + \, \bs{\gamma}-\oint_{T[1]\partial
  \Sigma_d}\,\dd^{d-1}\hat z \ \mathrm{ev}^* (\beta)
\ee
on $\mbf\cM_{\cL_{\beta}'}$, which is given by shifting the Liouville
potential $\vartheta$ to $\vartheta - \dd \beta$ with
$\cL_\beta'=\e^{\delta_\beta} \cL'$ the new zero locus of the shifted
Liouville potential. 

\subsection{Dimensional reduction by gauge fixing}
\label{sec:gaugefixing}

In AKSZ constructions the fields and antifields are not distinguished
from the onset. The theory is specified once the antifields are
assigned in the entire field content, and different choices yield
different field theories. In~\cite{Kokenyesi2018} we also studied the
gauge fixings of AKSZ theories from a different perspective.

In the usual BV quantized theories, the fields and antifields are
distinguished from the start, with the former including the physical
and ghost fields from the BRST picture, while the latter define canonically conjugate variables
with respect to the symplectic phase space structure on the space
of all fields $\mbf\cM$. 
Gauge fixing is then equivalent to a choice of a
Lagrangian submanifold $\mbf\cL$ of $\mbf\cM$. The Batalin-Vilkovisky theorem~\cite{Batalin1981}
ensures that the path integral over $\mbf\cL$ is independent of the
choice of representative for the homology class of the Lagrangian submanifold $\mbf\cL$. 
The Lagrangian submanifold intersects
the gauge orbits transversally, i.e. the action of the
BV--BRST charge $\bv{\bs{\cS}}{\, \cdot \,}$ vanishes on Lagrangian submanifolds,
as the BV bracket acts as zero there. Thus the BV gauge symmetry is
completely fixed on Lagrangian submanifolds.

In the following we introduce a particular gauge fixing as a
dimensional reduction technique which reduces a given AKSZ theory on the superworldvolume $\cW=T[1]\Sigma_d$ to an AKSZ theory on its boundary $\partial\cW=T[1]\partial\Sigma_d$.
We consider the case when the fields (but not the antifields) occur in even number and can be paired: a superfield $\bs{\phi}^a(\hat z)$ with ghost degree $|a|$ is paired with another superfield $\bs{\chi}_a(\hat z)$ with ghost degree $d-2-|a|$, and vice versa. The BV symplectic form is written in the canonical form 
\be \label{eq:AKSZgenBVsymplGF}
\bs{\omega} \, = \int_{{T}[1]\Sigma_d} \,\dd^d\hat z \ \big(
(-1)^{d+|a|} \, \bs{\delta}
\bs{\phi}^+_a(\hat z)\, \bs{\delta} \bs{\phi}^a(\hat z) \, + \, (-1)^{|a|}\, \bs{\delta}
\bs{\chi}^{a \, +}(\hat z)\, \bs{\delta} \bs{\chi}_a(\hat z) \big) \ , 
\ee
where we chose a convenient ordering of antifields $\bs{\phi}_a^+,\bs{\chi}^{a\,+}$ and
fields $\bs{\phi}^a,\bs{\chi}_a$ in this way. The ghost degrees of the antifields $\mbf\phi_a^+$ and $\bs{\chi}^{a \, +}$ are
$d-1-|a|$ and $|a|+1$ respectively. An arbitrary superfield $\mbf\phi^a$ can be expanded in terms of the degree~1 fiber coordinates $\theta^\mu$ of $\cW=T[1]\Sigma_d$ in the form
\be \label{eq:superfieldexp}
\bs{\phi}^a(\hat z) \, = \, \phi^{(0) \, a} (\sigma) \, + \,
\sum_{p=1}^d\, \frac{1}{p!}\, \phi^{(p) \, a}_{\mu_1 \cdots \mu_p} (\sigma) \, \theta^{\mu_1} \cdots \theta^{\mu_p} \ ,
\ee
where $\phi^{(p) \, a}$ are the degree $|a|-p$ coefficients of $\bs{\phi}^a$ which can be identified with $p$-forms on $\Sigma_d$.

We choose a submanifold as gauge fixing on the space of superfields
$\mbf\cM$. It is given by the constraints\footnote{It is important to
  note that the fields and antifields in this gauge are assigned in
  the bulk $\cW\setminus\partial\cW$.}
\be
\bs{\phi}^+_a \, = \, \bs{D} \bs{\chi}_a \qquad \mbox{and} \qquad   \bs{\chi}^{a \, +} \, = \, (-1)^{|a|\,(d+1)+1} \, \bs{D} \bs{\phi}^a \ , 
\ee
which reduces the BV symplectic form to
\be
\bs{\omega}_{\rm{gf}} \, = \oint_{{T}[1]\partial\Sigma_d} \,\dd^{d-1}\hat z \ (-1)^{d+|a|+1}\, \bs{\delta}
\bs{\chi}_a(\hat z)\, \bs{\delta} \bs{\phi}^a(\hat z) \ .
\ee
In the following we refer to this gauge as the \emph{exact gauge}. If
the worldvolume has no boundaries or the boundary conditions give
$\bs{\omega}_{\rm{gf}}=0$, the submanifold is a Lagrangian submanifold
as well, and hence it gives a full gauge fixing. Otherwise the
submanifold is not Lagrangian, and therefore it only gives a full
gauge fixing in the bulk $\cW\setminus\partial\cW$ but not on the
boundary $\partial\cW$.

If the Liouville potential is chosen as 
\be
\bs{\vartheta} \, = \, \int_{T[1]\Sigma_d}\, \dd^d \hat{z} \ \big(
(-1)^{d+|a|} \,
\bs{\phi}^+_a(\hat z)\,  \bs{\delta} \bs{\phi}^a(\hat z) \, + \, (-1)^{(d+1)\,|a|} \,
\bs{\chi}_a(\hat z)\, \bs{\delta} \bs{\chi}^{a \, +}(\hat z)  \big) \ ,
\ee
then the kinetic part of the AKSZ action is given by
\be
\bs{\cS}_0 \, = \, -\iota_{\bs{Q}_0}\bs{\vartheta} \, =  \, \int_{T[1]\Sigma_d}\, \dd^d \hat{z} \ \big( (-1)^{d+|a|+1} \,
\bs{\phi}^+_a(\hat z)\, \bs{D} \bs{\phi}^a(\hat z) \, + \,
(-1)^{(d+1)\,|a|+1} \,
\bs{\chi}_a (\hat z)\, \bs{D} \bs{\chi}^{a \, +}(\hat z)
\big) \ .
\ee
We have not specified any boundary conditions yet. They are needed in order to derive consistent equations of motion. The variation of the action $\bs{\cS}_0$ gives
\be \begin{aligned}
\delta\bs{\cS}_0 \, = \, \int_{T[1]\Sigma_d}\, \dd^d \hat{z} \ \big( & (-1)^{d+|a|+1} \,
\delta\bs{\phi}^+_a(\hat z)\, \bs{D} \bs{\phi}^a(\hat z) \, + \,
(-1)^{d+|a|+1} \, \bs{\phi}^+_a(\hat z)\, \bs{D} \delta \bs{\phi}^a(\hat z) \\
 & + \, (-1)^{(d+1)\,|a|+1} \,\bs{\chi}_a(\hat z)\, \bs{D}  \delta\bs{\chi}^{a \, +}(\hat z) \, + \, (-1)^{(d+1)\,|a|+1} \,
\bs{\chi}_a (\hat z)\, \bs{D} \delta \bs{\chi}^{a \, +}(\hat z)
\big) \ .
\end{aligned} \ee
The equations of motion for $\bs{\phi}^a$ and $\bs{\chi}_a$ are obtained via integration by parts. The boundary terms of the variation 
\be
\delta\bs{\cS}_0 \big|_{T[1]\partial\Sigma_d} \, = \, \oint_{T[1]\partial\Sigma_d} \,\dd^{d-1}\hat z \ \big(\bs{\phi}^+_a(\hat z)\, \delta \bs{\phi}^a(\hat z) \, - \, (-1)^{d\,(|a|+1)}\, \bs{\chi}_a(\hat z)\, \delta \bs{\chi}^{a \, +}(\hat z) \big)
\ee
must vanish on their own. The straightforward boundary conditions
$\bs{\phi}^+_a|_{T[1]\partial\Sigma_d}=0$, $\bs{\chi}^{a \,
  +}|_{T[1]\partial\Sigma_d}=0$ and
$\delta\bs{\phi}^a|_{T[1]\partial\Sigma_d}=0$,
$\delta\bs{\chi}_a|_{T[1]\partial\Sigma_d}=0$ result in a vanishing
reduced kinetic action on the boundary, so they are not suitable for
us. On the other hand, the boundary variation term $\delta\bs{\cS}_0
\big|_{T[1]\partial\Sigma_d}$ in the partial exact gauge fixing reduces to
\be \begin{aligned}
\delta\bs{\cS}_{0,\rm{gf}} \big|_{T[1]\partial\Sigma_d} \, &= \, \oint_{T[1]\partial\Sigma_d} \,\dd^{d-1}\hat z \ \big( \bs{D} \bs{\chi}_a(\hat z)\, \delta \bs{\phi}^a (\hat z) \, + \, (-1)^{d+|a|}\, \bs{\phi}^a (\hat z)\, \bs{D} \delta \bs{\chi}_a(\hat z) \big) \\[4pt]
& = \, \oint_{T[1]\partial\Sigma_d} \,\dd^{d-1}\hat z \ \bs{D}
\big(\bs{\chi}_a(\hat z)\, \delta \bs{\phi}^a (\hat z) \big) \\[4pt] \, &= \, 0 \ .
\end{aligned} \ee
As we see, the exact gauge is consistent with the necessary boundary
conditions, which means the equations of motion are well-defined in
this gauge, and the master equation also holds because the interaction
term reduces to the boundary as well. This is not true without the
exact gauge or suitable boundary conditions. Hence the exact gauge fixing appears here as a boundary condition.

The gauge fixed kinetic action
\be 
\bs{\cS}_{0, \rm{gf} }\, =  \, \oint_{T[1]\partial\Sigma_d}\, \dd^{d-1} \hat{z} \ (-1)^{d+|a|+1} \,
\bs{\chi}_a(\hat z)\, \bs{D} \bs{\phi}^a(\hat z)
\ee
can be derived from the Liouville potential 
\be
\bs{\vartheta}_{\rm{b}} \, = \, \oint_{T[1]\partial\Sigma_d}\, \dd^{d-1} \hat{z} \ (-1)^{d+|a|+1} \,
\bs{\chi}_a(\hat z)\, \bs{\delta} \bs{\phi}^a(\hat z) \ , 
\ee
with $\bs{\omega}_{\rm{gf}}=\mbf\delta \bs{\vartheta}_{\rm{b}}$, but
with the opposite sign:
\be
\bs{\cS}_{0, \rm{gf} }\, =  \, \iota_{\mbf Q{}_{0,\rm{b}}} \mbf \vartheta{}_{\rm{b}} \ ,
\ee
where $\mbf Q{}_{0,\rm{b}}$ is the cohomological vector field on $T[1]\partial\Sigma_d$.
The interaction term enters into the picture in a simpler way. Let us
assume that the Hamiltonian functional
$\mbf\gamma=\int_{T[1]\Sigma_d}\, \dd^d\hat z \ \rm{ev}^*(\gamma)$,
which satisfies the equation
$\iota_{\bs{Q}_\gamma}\bs{\omega}=\mbf\delta \mbf\gamma$, reduces to
the boundary in the exact gauge as
\be
\mbf\gamma{}_{\rm{gf}} \, = \, - \int_{T[1]\Sigma_d}  \, \dd^{d}\hat z
\ \mbf D \,  \rm{ev}^*(\beta) \, = \, - \oint_{T[1]\partial\Sigma_d}
\, \dd^{d-1}\hat z \ \rm{ev}^*(\beta) \, = : \, - \, \mbf \beta
\ee
for a function $\beta$ on the target graded manifold $\cM$ with degree
$d-1$. Then the full action \eqref{eq:AKSZactiongeneral} reduces in
the exact gauge to
\be
\bs{\cS}_{\rm{gf}} \, = \, \, \iota_{\mbf Q{}_{0,\rm{b}}} \mbf \vartheta{}_{\rm{b}} \, - \, \mbf \beta \ ,
\ee
which satisfies the BV master equation, and thus gives an AKSZ action on the boundary.

\subsection{Dimensional reduction by effective actions}
\label{sec:DimRedMeth}

In this paper we shall also apply another dimensional reduction
method, called `Losev's trick'~\cite{Mnev2008}, which is not specific
to boundary reductions. We briefly recall the technique
following~\cite{Cattaneo2009}, see also~\cite{Ikeda2007} where a
similar technique is employed. 

The symplectic structure $\omega$ on the target supermanifold $\cM$
induces a natural second order differential operator $\Delta$, which
in local coordinates is given by
\be 
\Delta=\frac12\,\omega^{\hat\imath\hat\jmath}\,
\frac{\rd}{\partial\hat X^{\hat\imath}} \, \frac{\rd}{\partial\hat X^{\hat\jmath}} \ ,
\ee
where $\omega^{\hat\imath\hat\jmath}$ is the inverse of
$\omega_{\hat\imath\hat\jmath}$. This pulls back to give the BV
Laplacian $\mbf\Delta$ for the BV bracket
$(\,\cdot\,,\,\cdot\,)_{\mathrm{BV}}$ on the space of AKSZ fields
$\mbf\cM$. The AKSZ action $\mbf \cS$ satisfies the BV quantum master
equation $\mbf\Delta\e^{-\bs{\cS}/\hbar}=0$ on $\mbf\cM$, which is
equivalent to $
\frac12\, (\mbf \cS,\mbf \cS)_{\mathrm{BV}} = \hbar \, \mbf\Delta\mbf \cS $. This ensures
independence of the BRST-invariant quantum field theory on the choice of
gauge fixing, provided we define the path integral
by equiping $\mbf\cM$ with
a measure $\mbf\mu$ which is compatible with $\mbf\omega$~\cite{Batalin1981}.

Borrowing standard terminology from renormalization of quantum field
theory, let us now assume that the space of AKSZ fields can be
decomposed into a direct product 
$
\mbf\cM=\mbf\cM_{\mathrm{UV}}\times\mbf\cM_{\mathrm{IR}}
$
of ultraviolet (UV) and infrared (IR) degrees of freedom, with a compatible decomposition of the canonical symplectic form $\mbf\omega=\mbf\omega_{\mathrm{UV}} + \mbf\omega_{\mathrm{IR}}$. Then the BV Laplacian also decomposes as $\mbf\Delta=\mbf\Delta_{\mathrm{UV}} + \mbf\Delta_{\mathrm{IR}}$. One now `integrates out' the ultraviolet degrees of freedom to get an effective action. The integration requires a gauge fixing on the ultraviolet sector $\mbf\cM_{\mathrm{UV}}$ of the space of superfields, which means a choice of a Lagrangian submanifold $\mbf\cL \subset \mbf\cM_{\mathrm{UV}}$. Then the effective BV action $\bs{\cS}_{\mathrm{eff}}$ in the infrared sector is defined as
\be
\e^{-\bs{\cS}_{\mathrm{eff}}/\hbar} \, := \, \int_{\mbf\cL}\, \sqrt{\mbf\mu}_{\mbf\cL} \ \e^{-\bs{\cS}/\hbar} \  ,
\ee
where $\sqrt{\mbf\mu}_{\mbf\cL}$ is the measure on $\mbf\cL$ induced by $\mbf\mu$. Therefore the effective action satisfies the quantum master equation $\mbf\Delta_{\mathrm{IR}}\e^{-\bs{\cS}_{\mathrm{eff}}/\hbar}=0$. A change of gauge fixing in the ultraviolet sector only changes $\e^{-\bs{\cS}_{\mathrm{eff}}/\hbar}$ by a $\mbf\Delta_{\rm{IR}}$-exact term. Similarly, the value of the partition function is independent of the particular choice of splitting $\mbf\cM=\mbf\cM_{\mathrm{UV}}\times\mbf\cM_{\mathrm{IR}}$ by the Batalin-Vilkovisky theorem~\cite{Batalin1981}.
In the following we use this method to reduce three-dimensional AKSZ
sigma-models to AKSZ sigma-models in two dimensions.

\section{String sigma-models}
\label{sec:StringAKSZ}

In this section we describe several relevant examples of
two-dimensional AKSZ sigma-models which are related to the topological A- and B-models.

\subsection{Poisson sigma-model}
\label{sec:AKSZPoisson}

In dimension $d=2$, the AKSZ theory with target space a degree~1
QP-manifold describes the topological sigma-model for strings
in an NS--NS $B$-field background, whose first order formalism is the
Poisson sigma-model~\cite{Ikeda1993,Schaller1994} whereby BV quantization yields the Cattaneo-Felder 
path integral approach~\cite{CattFeld1999}. 
The AKSZ formulation of the Poisson sigma-model is studied in~\cite{Cattaneo2001}.
We take $\cW= T [1]\Sigma_2$ for an oriented Riemann surface $\Sigma_2$, and $\cM= T ^*[1]M$ with degree~0 base
coordinates $X^i$ on the target space $M$ and degree~1 fiber coordinates $\chi_i$. The canonical symplectic form on $\cM$ is 
\be
\omega_2=\dd \chi_i \wedge \dd X^i \  ,
\ee
which leads to the canonical graded Poisson bracket $\{ X^i, \chi_j \} = \delta^i{}_j$ on the local coordinates of $\cM$.
We choose the Liouville potential to be $\vartheta_2= \chi_i\, \dd
X^i$. Its zero locus is $\cL_2=\{ \chi_i=0 \}$. The most general form of a degree~2 Hamiltonian function
$\gamma_\pi$ on $\cM$ is given by a $(0,2)$-tensor $\pi$ on $M$ as
\be
\gamma_\pi \, = \, \frac 12 \, \pi^{ij}(X) \, \chi_i \, \chi_j \ .
\ee 
Compatibility of the corresponding cohomological vector field $Q_{\gamma_\pi}$ with $\omega_2$ and 
the classical master equation $\{ \gamma_\pi, \gamma_\pi \}=0$ implies
that $\pi$ must be a Poisson bivector on $M$, i.e. $[\pi,\pi]_{\rm S}=
0$, where $[\,\cdot\,,\,\cdot\,]_{\rm S}$ denotes the Schouten bracket
on multivectors. In other words, a QP1-manifold is the same thing as a Poisson manifold $(M,\pi)$, which by construction is also a Lie algebroid on the cotangent bundle $T^*M$. 
The Hamiltonian function determines a derived bracket which defines a Poisson bracket on $C^\infty(M)$ through
\be
\{ f,g \}_{\pi} \, = \, \pi(\dd f\wedge\dd g) \, = \, - \, \{ \{ f , \gamma \} , g \} \  .
\ee

The kinetic part of the AKSZ action is inherited from the cohomological vector field $Q_{\cW}$ on $ \cW=T [1]\Sigma_2$, and is given by
\be
\bs{\cS}_{0}^{(2)}\, = \, \int_{ T [1]\Sigma_2} \,\dd^2\hat z \ \bs{\chi}_i \, \bs{D}  \bs{X}^i \ ,
\ee
where as before the superworldsheet differential is $\bs{D}=\theta^\mu\, \frac\partial{\partial\sigma^\mu}=Q_\cW$. 
Together these ingredients give the AKSZ action for the Poisson sigma-model as
\be \label{eq:PoissonActAKSZ}
\bs{\cS}_{\pi}^{(2)} \, = \, \int_{ T [1]\Sigma_2}\,\dd^2\hat z \ \Big(\, \bs{\chi}_i \, \bs{D}  \bs{X}^i \, + \, \frac 12 \, \bs{\pi}^{ij} \, \bs{\chi}_i\, \bs{\chi}_j \Big) \ ,
\ee
where $\mbf f=\mbf\phi^*(f)=f(\mbf\phi)$ for a function $f$ on $\cM$
and $\mbf\phi\in\mbf\cM$.
Integrating over the odd coordinates $\theta^\mu$ and restricting to
the degree~0 fields in \eqref{eq:PoissonActAKSZ} yields the
first order string sigma-model action 
\be\label{eq:Poissonaction}
 \int_{\Sigma_2}\, \Big( \chi_i \wedge \dd X^i \, + \, \frac 12 \, \pi^{ij} \,  \chi_i \wedge \chi_j \Big) \ ,
 \ee
and if $\pi$ is non-degenerate it is classically equivalent to the
topological bosonic string $B$-field coupling
\be
\int_{\Sigma_2}\, X^*(B) \, = \, \frac 12\, \int_{\Sigma_2}\, B_{ij} \, \dd X^i \wedge \dd X^j \ ,
\ee
where the flat Kalb-Ramond two-form field $B$ on $M$ is the inverse of the bivector $\pi$.

\subsection{AKSZ formulations of the A-model}

The topological A- and B-models coupled to gravity are the topological
A- and B-model string theories, which have been widely studied for
more than 20 years. They were also one of the first examples of the AKSZ
construction in~\cite{AKSZ1997}. In the following we review their
relevant AKSZ constructions, which reduce to the A- or B-model in a
particular gauge. The reader can find details about their
field-antifield choices and gauge fixing in the indicated references,
and therefore we only define their AKSZ sigma-models.

We begin with the A-model, whose AKSZ constructions were mostly
related to the Poisson sigma-model or the $B$-field
coupling. Recently a different approach related to the AKSZ sigma-model of
topological membranes on $G_2$-manifolds appeared in~\cite{Kokenyesi2018}.

\medskip

{\underline{\sl A1.} \ } 
The original AKSZ construction~\cite{AKSZ1997} is formulated in the
same way as the Poisson sigma-model in \S\ref{sec:AKSZPoisson} but
with zero kinetic term. Thus it has the same target QP1-manifold
with the same symplectic structure and Hamiltonian as those of the
Poisson sigma-model, where the Poisson bivector $\pi$ is given by the
inverse of the K{\"a}hler form on the target Calabi-Yau manifold. The
AKSZ action thus constructed is 
\be \label{eq:AmodelAKSZ1997}
\bs{\cS}_{\rm{A}1}^{(2)} \, = \, \frac 12 \, \int_{T[1]\Sigma_2}  \, \dd^2 \hat z \  \bs{\pi}^{ij}\, \bs{\chi}_i\, \bs{\chi}_j \ .
\ee

\medskip

{\underline{\sl A2.} \ }
A complete Poisson sigma-model formulation for the A-model with
kinetic term \eqref{eq:PoissonActAKSZ} appeared in~\cite{Pestun2006} as
\be \label{eq:AmodPoissonAKSZaction}
\bs{\cS}_{\rm{A}2}^{(2)} \, = \, \bs{\cS}_{\pi}^{(2)} \, = \, \int_{T[1]\Sigma_2}  \, \dd^2 \hat z \  \Big(\, \bs{\chi}_i\, \bs{D}  \bs{X}^i \, + \, \frac 12 \, \bs{\pi}^{ij}\, \bs{\chi}_i\, \bs{\chi}_j \Big) \ .
\ee
The equation of motion for $\bs{\chi}_i$ reduces it to the AKSZ action
\eqref{eq:AmodelAKSZ1997} up to a sign, so they are classically equivalent.

\medskip

{\underline{\sl A3.} \ }
The BV quantized topological NS--NS $B$-field coupling is not
strictly speaking constructed by the AKSZ formalism, but it is
nevertheless worth mentioning as a BV action which gives the
A-model~\cite{Zucchini2004} with the same field definitions as those of the Poisson sigma-model:
\be
\bs{\cS}_{\rm{A}3}^{(2)} \, = \, \frac12 \, \int_{T[1]\Sigma_2}  \, \dd^2 \hat z \  \bs{B}_{ij} \, \bs{D}\bs{X}^i\, \bs{D}\bs{X}^j \ ,
\ee
where the two-form $B$ is the  K{\"a}hler form, and the flat condition $\dd
B=0$ is equivalent to the Poisson condition of its inverse
$\pi$. It is of course not surprising
 that the $B$-field coupling is classically equivalent to the Poisson sigma-model as well.

\medskip

{\underline{\sl A4.} \ }
 In~\cite{Zucchini2004,Stojevic2005} an AKSZ Poisson sigma-model
 together with the topological $B$-field coupling is used as an AKSZ
 formulation of the A-model with action
 \be 
\bs{\cS}_{\rm{A}4}^{(2)} \, = \, \int_{T[1]\Sigma_2}  \, \dd^2 \hat z
\  \Big(\, \bs{\chi}_i\, \bs{D} \bs{X}^i \, + \, \frac 12 \,
\bs{\pi}^{ij}\, \bs{\chi}_i\, \bs{\chi}_j \, +  \, \frac 14 \, \bs{B}_{ij} \, \bs{D}\bs{X}^i\, \bs{D}\bs{X}^j \Big) \ ,
\ee
where $B_{ij}$ is the inverse of $\pi^{ij}$. The last term has no effect in the BV bracket since $\dd B = 0$. 

\medskip

{\underline{\sl A5.} \ }
Another simple construction also gives the A-model~\cite{Kokenyesi2018}. Let the target supermanifold be $\cM=T^*[1]T[1]M$ where the local coordinates are given by $(X^i,\chi_i,\zeta^i,y_i)$ with degrees $(0,1,1,0)$. The symplectic form is
\be
\omega_{\rm{A}5} \, = \, \dd \chi_i \w \dd X^i \, + \, \dd y_i \w \dd \zeta^i \ .
\ee 
The Hamiltonian
\be
\gamma_{\rm{A}5} \, = \, \chi_i \, \zeta^i
\ee
gives an AKSZ action with zero kinetic term as
\be
\bs{\cS}_{\rm{A}5}^{(2)} \, = \, \int_{T[1]\Sigma_2}  \, \dd^2 \hat z
\  \bs{\chi}_i \, \bs{\zeta}^i \ .
\ee
An interesting feature of this AKSZ construction is that it can be
reproduced from an AKSZ membrane theory, namely the standard Courant
sigma-model which we discuss in \S\ref{sec:CourantSigmaMod}. The
reduction can be performed by taking all fields to be independent of
the extra coordinate direction.

\medskip

{\underline{\sl A6.} \ } 
A somewhat different construction was proposed in~\cite{Kokenyesi2018}, which uses degree $-1$ target space coordinates. Thus the target space is not strictly speaking a QP-manifold anymore, and this takes us out of the realm of graded geometry into derived geometry, but its AKSZ construction is still applicable wherein using negative degree fields such as ghosts and antifields is natural.
The target derived manifold is $\cM=T^*[1]T[-1]T[1]M$ on which the local coordinates of $T[-1]T[1]M$ are $(X^i,\zeta^i,b^i,\eta^i)$ with degrees $(0,1,0,-1)$, and $X^i$ associated to $M$. The cotangent fiber coordinates are $(\chi_i, y_i, n_i, h_i)$ with degrees $(1,0,1,2)$, and the symplectic structure is 
\be
\omega_{\rm{A}6} \, = \, \dd \chi_i \w \dd X^i \, + \, \dd y_i \w \dd \zeta^i \, + \, \dd h_i \w \dd \eta^i \, + \, \dd n_i \w \dd b^i \ .
\ee
The Hamiltonian
\be
\gamma_{\rm{A}6} \, = \, \chi_i \, \zeta^i \, + \, b^i \, h_i
\ee
gives the AKSZ action
\be
\bs{\cS}_{\rm{A}6}^{(2)} \, = \, \int_{T[1]\Sigma_2}  \, \dd^2 \hat z \ \big(\bs{\chi}_i \, \bs{\zeta}^i \, + \, \bs{b}^i \, \bs{h}_i\big) 
\ee
with zero kinetic term. This construction can be obtained from a similar reduction of the standard Courant sigma-model as in the previous construction.
These latter two AKSZ constructions can be obtained by dimensional reductions of AKSZ built topological membrane theories on $G_2$-manifolds, which can be reformulated as an AKSZ threebrane theory related to exceptional generalized geometry of M-theory~\cite{Kokenyesi2018}. 

\medskip

{\underline{\sl Zucchini model.} \ }
The BV sigma-model of~\cite{Zucchini2004} is not strictly speaking given by an AKSZ construction, since it involves BV quantized kinetic terms which do not arise from a Louville potential of the BV symplectic form. It has the same field content and BV symplectic form as those of the Poisson sigma-model, and the BV action is given by
\be \label{eq:ZucchiniAction}
\bs{\cS}_{\rm{Z}}^{(2)} \, = \, \int_{T[1]\Sigma_2} \, \dd^2 \hat z \ \Big(\, \bs{\chi}_i \, \bs{D}  \bs{X}^i \, + \, \frac 12 \, \bs{\pi}^{ij} \, \bs{\chi}_i \, \bs{\chi}_j \, + \, \frac 12 \, \bs{\omega}_{ij} \, \bs{D} \bs{X}^i \, \bs{D
} \bs{X}^j \, + \, {\bs{J}^i}_j \, \bs{\chi}_i \, \bs{D} \bs{X}^j \Big) \ ,
\ee
where $\pi$ is a bivector and $\omega$ is a two-form, and together with the $(1,1)$-tensor $J$ they satisfy the identities 
\be \begin{aligned}
 {J^i}_k \, {J^k}_j \, + \, \pi^{ik} \, \omega_{kj} \, + \, \delta^i{}_j \, &= \, 0 \ , \\[4pt]
 {J^i}_k \, \pi^{kj} \, + \, {J^j}_k \, \pi^{ki} \, &= \, 0 \ , \\[4pt]
 \omega_{ik} \, {J^k}_j \, + \, \omega_{jk} \, {J^k}_i \, &= \, 0 \ . 
\end{aligned}\ee
The master equation imposes further constraints
\be \begin{aligned} \label{eq:GenComplexStrId}
 \pi^{\left[i\right|l} \, \partial_l \pi^{\left.jk\right]} \, &= \, 0 \ , \\[4pt]
 {J^l}_i \, \partial_l \pi^{jk} \, + \, 2 \, \pi^{jl} \, \partial_{[i} {J^k}_{l]} \, + \, \pi^{kl} \, \partial_l {J^j}_i \, - \, {J^j}_l \, \partial_i \pi^{lk} \, &= \,  0 \ , \\[4pt]
 2 \, {J^l}_{[i|} \, \partial_l {J^k}_{|j]} \, - \, 2 \, {J^k}_{l} \, \partial_{[i} {J^{l}}_{j]} \, + \, 3 \, \pi^{kl} \, \partial_{[l} \omega_{ij]} \, &= \, 0 \ , \\[4pt]
 {J^l}_i \, \partial_{[l} \omega_{jk]} \, + \, {J^l}_j \, \partial_{[l} \omega_{ki]} \, + \, {J^l}_k \, \partial_{[l} \omega_{ij]} \, - \, \partial_{[i} \big(\omega_{j|l} \, {J^l}_{|k]} \big) \, &= \, 0 \ ,
\end{aligned}\ee
where $\partial_i= \partial / \partial X^i$, which are the same identities as the integrability condition of a generalized complex structure $\mathbb{J}$ in the form
\be \label{eq:GenComplexStrMatrix} 
{\mathbb{J}^I}_J \, = \, \begin{pmatrix} {J^i}_j & \pi^{ij} \\ \omega_{ij} & -{J^j}_i \end{pmatrix} \ ,
\ee
where the doubled indices $I,J$ have been introduced. The Zucchini model reduces to the Poisson sigma-model upon setting $J=0$ and $\omega=0$, which is the A-model. If in addition $\omega$ is non-zero it adds a $B$-field coupling, which is just another copy of the A-model.

\subsection{AKSZ formulations of the B-model}

AKSZ constructions for the topological B-model are more diverse and have different superfield contents. We do not enumerate all of them here, nor the original construction from~\cite{AKSZ1997}, since they are similar to the ones described below.

\medskip

{\underline{\sl B1.} \ } 
The base degree 0 manifold of the target QP-manifold, which is a Calabi-Yau threefold $M$, is equiped with a complex structure which splits the local coordinate indices to $i=(a,\overline{a})$, where $a=1,2,3$. The target QP-manifold $\cM$ is defined by its coordinates: $X^a$, $X^{\overline{a}}$, $\tx_{\overline{a}}$ have degree 0, and $\chi_a$, $\chi_{\overline{a}}$, $\wt{\chi}{}^{\,\overline{a}}$ have degree 1. The symplectic form on $\cM$ is 
\be \label{eq:BmodSympHofman}
\omega_{\rm{B}1} \, = \, \dd X^a \w \dd \chi_a \, + \, \dd X^{\overline{a}} \w \dd \chi_{\overline{a}} \, + \, \dd \tx_{\overline{a}} \w \dd  \wt{\chi}{}^{\,\overline{a}} \ .
\ee 
The B-model is constructed in~\cite{Hofman2002b} by the AKSZ action
\be \label{eq:BmodAKSZHofman}
\bs{\cS}_{\rm{B}1}'^{\,(2)} \, = \, \int_{T[1]\Sigma_2} \, \dd^2 \hat z \ \big(\,\bs{\chi}_a \, \bs{D}\bs{X}^a \, + \, \bs{\chi}_{\overline{a}} \, \bs{D}\bs{X}^{\overline{a}} \, + \, \mbf \tx{}_{\overline{a}} \, \bs{D} \wt{\bs{\chi}}{}^{\,\overline{a}} \, + \, \bs{\chi}_{\overline{a}} \, \wt{\bs{\chi}}{}^{\,\overline{a}}\, \big) \ .
\ee

We can enlarge its field content with the addition of new coordinates
$\wt{\chi}{}^{\,a}$ and $\tx{}_a$ whose contribution to the symplectic
structure is defined by the term
\be \label{eq:BmodSympHofmanEnlarged}
\dd \tx_a  \w \dd \wt{\chi}{}^{\,a} \ ,
\ee
and furthermore we also add the term $\mbf\tx{}_a \, \bs{D}
\wt{\bs{\chi}}{}^a + \bs{\chi}_a \, \wt{\bs{\chi}}{}^a$ to the AKSZ
action \eqref{eq:BmodAKSZHofman} which can be set to zero with gauge
fixing $\wt{\bs{\chi}}{}^a=0$. Introducing the new fields leads to an
extended AKSZ action for the B-model given by
\be \label{eq:BmodAKSZaction1}
\bs{\cS}_{\rm{B}1}^{(2)} \, = \, \int_{T[1]\Sigma_2}  \, \dd^2 \hat z
\ \big(\bs{\chi}_i \, \bs{D}\bs{X}^i \, + \, \mbf \tx{}_{i} \, \bs{D}
\wt{\bs{\chi}}{}^{i} \, + \, \bs{\chi}_{i} \, \wt{\bs{\chi}}{}^{i} \big) \ .
\ee

\medskip

{\underline{\sl B2.} \ } 
The AKSZ construction of the B-model with an explicit complex
structure $J$ was studied in~\cite{Ikeda2007}, see also~\cite{Ikeda2012}. It has the same field content as the first construction of the B-model: $X^i$, $\tx_i$ are degree 0 coordinates and $\chi_i$, $\zeta^i$ are degree 1 coordinates. The symplectic structure only differs in a sign from the first construction:
\be \label{eq:BmodAKSZ3}
\omega_{\rm{B}2} \, = \, \dd X^i \w \dd \chi_i \, - \, \dd \tx_i \w \dd \wt{\chi}{}^{\,i} \ .
\ee
The AKSZ action is given by
\be \label{eq:BmodAKSZaction3}
\bs{\cS}_{\rm{B}2}^{(2)} \, = \, \int_{T[1]\Sigma_2}  \, \dd^2 \hat z
\ \big(\bs{\chi}_i \, \bs{D}\bs{X}^i \, - \, \mbf \tx{}_{i} \, \bs{D}
\wt{\bs{\chi}}{}^{i} \, + \, {J^i}_j \, \bs{\chi}_{i} \, \wt{\bs{\chi}}{}^{j} \big) \ ,
\ee
where ${J^i}_j$ is a constant complex structure on the target
manifold. The first construction is just a special case of this: If we
take $J^i{}_j = \ii \delta^i{}_j$, and rescale the fields $\mbf
\tx{}_i$ and $\wt{\bs{\chi}}{}^i$ by $\ii$, then the action
\eqref{eq:BmodAKSZaction3} reduces to $\bs{\cS}_{\rm{B}1}^{(2)}$.

The case of non-constant complex structure $J$ was also studied
in~\cite{Ikeda2007}, and an AKSZ sigma-model was proposed, of which the master equation gives the integrability condition 
\be \label{eq:ComplStrComp}
{J^l}_{[i|} \,\partial_{{l}} {J^k}_{|j]}  - \, {J^k}_l \,\partial_{[i} {J^l}_{j]} \, = \, 0 \ ,
\ee
and the condition
\be
{J^i}_k \, {J^k}_j \, = \, - \, \delta^i{}_j
\ee
is added by hand. The field content is the same as that of the constant case, and 
the action constructed by the AKSZ formalism is given by
\be \label{eq:ComplexStrAKSZaction}
\bs{\cS}_{J}^{(2)} \, = \, \int_{T[1]\Sigma_2} \, \dd^2 \hat z \
\big(\bs{\chi}_i \, \bs{D}\bs{X}^i \, - \, \mbf \tx{}_{i} \, \bs{D}
\wt{\bs{\chi}}{}^{i} \, + \, {\mbf J^i}_j \, \bs{\chi}_{i} \,
\wt{\bs{\chi}}{}^j \, + \,\bs{\partial}_j \bs{J}{}^i{}_k \, \mbf
\tx{}_i \, \wt{\bs{\chi}}{}^j \, \wt{\bs{\chi}}{}^k \big) \ .
\ee

\section{Courant sigma-models}
\label{sec:CourantSigmaMod}

In this section we review the three-dimensional Courant sigma-model,
and its specific examples which are relevant for us: the standard
and the contravariant Courant sigma-models. We also study their boundary
reductions in the exact gauge.

\subsection{Courant algebroids}
\label{sec:CourantAlgAKSZ}

In dimension $d=3$, the corresponding AKSZ sigma-model with source
dg-manifold $\cW=T[1]\Sigma_3$ is defined on
membrane worldvolume superfields
with target space a QP-manifold of degree~2, which corresponds to a
Courant algebroid~\cite{Roytenberg2002b}. 
In this paper we work only with the QP2-manifold $\cM = T^*[2]T[1]M$. We choose local Darboux
coordinates $(X^i,\psi^i,\chi_i,F_i)$ with degrees $(0,1,1,2)$ in
which the graded symplectic structure is given by
\be \label{eq:StdCourantSympl} 
\omega_3 \, = \, \dd X^i \w \dd F_i \, + \, \dd \chi_i \w \dd \psi^i \ .
\ee
The graded Poisson brackets of the
coordinates are canonical in the sense that
$\{ X^i , F_j \} =\delta^i{}_j$ and $\{ \chi_i,\psi^j \} =
\delta_i{}^j$. For the Liouville potential we choose $\vartheta_3=F_i\, \dd
X^i-\chi_i\, \dd\psi^i$. Its zero locus is $\cL_3=\{F_i=0, \,
\psi^i=0\}$. The most general form of
the degree~3 Hamiltonian function is given by
\be \label{eq:HamiltonianCourantAlg} 
\gamma_{\rho,T} \, = \, \rho^i{}_I (X) \, F_i \, \zeta^I \, + \, \frac{1}{3!} \,
T_{IJK}(X) \, \zeta^I \, \zeta^J \, \zeta^K \ ,
\ee
for degree~0 functions $\rho^i{}_I$ and $T_{IJK}$ on $M$, where we introduced a
doubled index notation for the local degree~1 coordinates
$\zeta^I=(\psi^i,\chi_i)$. The three-form $T_{IJK}$ encodes the
allowed geometric and non-geometric supergravity fluxes for given $\rho^i{}_I$.

We now define three operations given by taking
derived brackets defined by $\gamma_{\rho,T}$ 
and the graded Poisson bracket through
 \be \label{eq:derivedStrCourantAlg}
[e_1,e_2]_{\rm D} =  \{\{ e_1 , \gamma_{\rho,T} \} , e_2 \} \ , \qquad
\langle e_1, e_2\rangle = \{ e_1,e_2 \} \qquad \mbox{and} \qquad
\rho(e) =  \{ e , \{ \gamma_{\rho,T} , \, \cdot \, \}\} \ .
\ee
These operations are defined on degree~1 functions $e$ with local
expression $e=f_I(X)\,\zeta^I$, where $f_I$ is a degree~0
function on the body $M=\cM_0$ of $\cM$, which are identified as local
sections of the generalized tangent bundle
\be
E= T M\oplus T ^*M
\ee
over $M$; symbolically
\be \label{eq:QPmanGeoCorrespondence}
 A^i \,\chi_i \, + \, \alpha_i \, \psi^i \ \longleftrightarrow \ A^i
 \,\frac\partial{\partial X^i} \, + \, \alpha_i \, \dd X^i \ .
\ee
The classical master equation $\{\gamma_{\rho,T},\gamma_{\rho,T}
\}=0$ then implies that they endow $E$ with the structure of a Courant
algebroid: A \emph{Courant algebroid} on a manifold $M$ is a vector bundle $E$ over $M$ equiped
with a symmetric non-degenerate bilinear form $\langle \, \cdot \, ,
\, \cdot \, \rangle$ on its fibers, an anchor map $\rho: E \rightarrow
T M$, and a binary bracket of sections $[\, \cdot \, , \, \cdot \,
]_{\rm D}$, called the Dorfman bracket, which together satisfy
\be\label{eq:CourantAx}
\begin{aligned}
{}[ e_1 ,[e_2,e_3 ]_{\rm D} ]_{\rm D} \, &= \ [[e_1,e_2]_{\rm D},e_3]_{\rm D} \, + \, [e_2,[e_1,e_3]_{\rm D}]_{\rm D} \  , \\[4pt]
\rho(e_1) \langle e_2,e_3 \rangle \, &= \ \langle [e_1,e_2]_{\rm D} , e_3 \rangle \, + \, \langle e_2 , [e_1,e_3]_{\rm D} \rangle \ , \\[4pt]
\rho(e_1) \langle e_2,e_3 \rangle \, &= \ \langle e_1 , [e_2, e_3 ]_{\rm D} \, + \, [e_3 , e_2]_{\rm D} \rangle \ , 
\end{aligned}
\ee
where $e_1,e_2,e_3$ are sections of $E$. In this paper we consider two
particular examples of Courant algebroids.

\medskip

{\underline{\sl Standard Courant algebroid.} \ }
The simplest Hamiltonian function with $\rho^i{}_I = \delta^i{}_I$ and
$T_{IJK}=0$ is given by
\be \label{eq:HamStdCourant}
\gamma_0 \, = \, F_i \, \psi^i \ .
\ee
Its derived brackets on degree 1 functions
\eqref{eq:QPmanGeoCorrespondence} gives the \emph{standard Courant algebroid} on the
generalized tangent bundle $E= T M \oplus  T ^*M$, which features in generalized geometry~\cite{Hitchin2004,Gualtieri2003}. It is an extension of the Lie
algebroid of tangent vectors by cotangent vectors 
with the three operations
   \be \begin{aligned}
 \langle A + \alpha , B +  \beta \rangle \, &= \ \iota_A\beta \, + \, \iota_B \alpha \ , \\[4pt]
 \rho(A + \alpha) \, &= \ A \ , \\[4pt]
 [A+\alpha,B+\beta]_{{\rm D},0}\, &= \ [A,B] \, + \, \cL_A \beta \, - \, \iota_B\, \dd \alpha \ ,  \label{eq:DorfmanBr}
 \end{aligned} \ee
where the sections of $E= T M\oplus T ^*M$ are composed of vector
fields $A,B$ and one-forms $\alpha,\beta$. The antisymmetrization of
the standard Dorfman bracket in \eqref{eq:DorfmanBr}, called the
Courant bracket, is the natural bracket in generalized geometry
which is compatible with the commutator
algebra 
of generalized Lie derivatives~\cite{Hitchin2004,Gualtieri2003}. 

Only
the simplest case of pure NS--NS flux $T_{IJK}=H_{ijk}$ is consistent with the choice of
anchor map $\rho^i{}_I = \delta^i{}_I$ of the standard Courant
algebroid, which is necessarily closed by the classical master equation. Given a Kalb-Ramond two-form field $B$ on $M$, with $H=\dd
B$, canonical transformation of the Hamiltonian function
\eqref{eq:HamStdCourant} by the degree~2 function $B=\frac12\,
B_{ij}(X)\, \psi^i\,\psi^j$ on $\cM$ yields the twisted Hamiltonian function
\be
\gamma_H \, := \, \e^{\delta_B}\gamma_0 \, = \, F_i \, \psi^i \, + \, \frac{1}{3!} \, H_{ijk} \,
\psi^i \, \psi^j \, \psi^k \ .
\ee
The NS--NS $H$-flux thus appears as a twisting of the standard Courant
algebroid, which gives rise to a deformation of the Dorfman bracket
through an extra term as
 \be \label{eq:CourantBrTwistedH}
 [A+\alpha,B+\beta]_{{\rm D},H}\, = \ [A,B] \, + \, \cL_A \beta \, -
 \, \iota_B\, \dd \alpha \, + \, \iota_A \iota_B H \ .
\ee

\medskip

{\underline{\sl Poisson Courant algebroid.} \ }
Consider the Hamiltonian defined through a bivector $\pi$ and a three-vector
$R$ on $M$ by setting $\rho^i{}_I=\pi^{ij}$, $T_{IJK}=(\partial_i\pi^{jk},R^{ijk})$ to give
\be \label{HamPoisson3D}
\gamma_{\pi,R} \, = \, \pi^{ij}\, F_i\, \chi_j \, - \, \frac 12\, \partial_i \pi^{jk}\, \psi^i\, \chi_j\, \chi_k \, + \, \frac{1}{3!} \, R^{ijk}(X)\, \chi_i\, \chi_j\, \chi_k \ .
\ee
The master equation $\{ \gamma_{\pi,R}, \gamma_{\pi,R} \}=0$ gives the constraints
\be \label{eq:ContraCourantMasterConstr}
[\pi,\pi]_{\rm{S}} \, = \, 0 \qquad \mbox{and} \qquad [\pi,R]_{\rm{S}} \, = \, 0 \ .
\ee
Note that the $R$-flux also enters here as a twist: The Hamiltonian \eqref{HamPoisson3D} can be regarded as
a canonical transformation
$\gamma_{\pi,R}=\e^{\delta_\beta}\gamma_{\pi,0}$ by a degree~2
function $\beta=\frac12\, \beta^{ij}(X)\, \chi_i\, \chi_j$ with
$R=\dd_\beta\beta$ where $\dd_\beta=[\beta,\, \cdot \,]_{\rm S}$, regarded
as a bivector $\beta$ on $M$ which is T-dual to the $B$-field of the
$H$-flux frame.
The corresponding Courant algebroid is the Poisson Courant algebroid~\cite{Asakawa2014}, for which the
identities are equivalent to the Poisson condition for $\pi$ if $R=0$: 
The \emph{Poisson Courant algebroid} is the Courant
algebroid on the generalized tangent bundle $E=TM \oplus T^* M$ over a Poisson manifold $(M,\pi)$ with the operations
 \be \begin{aligned}
 \langle A + \alpha , B +  \beta \rangle \, &= \ \iota_A\beta \, + \, \iota_B \alpha \ , \\[4pt]
 \rho(A + \alpha) \, &= \ \iota_\alpha\pi \ , \\[4pt]
 [A+\alpha,B+\beta]_{{\rm D};\pi,R}\, &= \ [\alpha,\beta]_{\pi} \, + \, \cL_\alpha^\pi Y \, - \, \iota_\beta\, \dd_\pi X \, - \, \iota_\alpha \iota_\beta R \ , 
 \end{aligned} \ee
where $\cL_\alpha^\pi=\iota_\alpha\, \dd_\pi + \dd_\pi\, \iota_\alpha$ and 
$[\, \cdot \, , \, \cdot \,]_\pi$ is the Koszul bracket on one-forms
given by
\be
[\alpha,\beta]_\pi \, = \, \cL_{\iota_\alpha\pi}\beta \, - \, \cL_{\iota_\beta\pi} \alpha \, - \, \dd \big(\pi(\alpha\wedge\beta)\big) \ .
\ee

\subsection{Standard Courant sigma-model}
\label{sec:StdCourantAlgAKSZ}

It is evident from the general construction that Courant algebroids
are uniquely encoded (up to isomorphism) in the corresponding AKSZ topological membrane theories, which are called Courant sigma-models~\cite{Roytenberg2007}. In the particular example of the standard Courant algebroid on $E=TM\oplus T^*M$ twisted by a closed NS--NS three-form flux $H$, the mapping space $\mbf\cM$ of superfields supports the canonical BV symplectic structure 
\be \label{eq:StdCourantBVomega}
\mbf\omega_3 = \int_{T[1]\Sigma_3}\,\dd^3\hat z \ \big(\mbf\delta\mbf
X^i\, \mbf\delta\mbf F_i + \mbf\delta\mbf\psi^i\, \mbf\delta\mbf\chi_i\big) \ ,
\ee
where the ghost number $2$ superfields $\mbf F_i$ and ghost number $0$ superfields $\mbf X^i$, as well as the conjugate pairs of
ghost number $1$ superfields $\mbf\chi_i$ and $\mbf\psi^i$, contain each other's antifields respectively. The AKSZ construction leads to the action
\be \label{eq:AKSZactionStdCourant}
\mbf \cS{}_{H}^{(3)} = \int_{T[1]\Sigma_3}\,\dd^3\hat z \ \Big(\mbf F_i\, \mbf D\mbf X^i -
\mbf\chi_i\, \mbf D\mbf\psi^i + \mbf F_i\, \mbf\psi^i + \frac1{3!}\, \mbf H_{ijk}\, \mbf\psi^i\,\mbf\psi^j\,\mbf\psi^k\Big) \ ,
\ee
which solves the classical master equation $\big(\mbf
\cS{}_{H}^{(3)},\mbf\cS{}_{H}^{(3)}\big)_{{\rm{BV}}}=0$. Integrating over
$\theta^\mu$ and restricting to degree~0 fields in
\eqref{eq:AKSZactionStdCourant} yields the first order membrane sigma-model action
\be \label{eq:Membr1}
\int_{\Sigma_3}\,\Big( F_i \w \big(\dd X^i - \psi^i \big) \, - \, \chi_i \w \dd \psi^i \, + \, \frac{1}{3!} \, H_{ijk} \, \psi^i \w \psi^j \w \psi^k \Big) \ ,
\ee
which is classically equivalent to the topological bosonic membrane Wess-Zumino coupling
\be\label{eq:3H}
\int_{\Sigma_3}\, X^*(H) \, = \, \frac{1}{3!}\,\int_{\Sigma_3}\, H_{ijk} \, \dd X^i \w \dd X^j \w \dd X^k \ .
\ee

The standard Courant sigma-model on an open worldvolume is well-defined
if, as usual, one specifies its boundary conditions. Instead we
consider it in exact gauge as an illustration. The exact gauge defined
in \S\ref{sec:gaugefixing} reads here as
\be \label{eq:ExactGaugeStdCourant}
\bs{F}_i \, = \, \bs{D} \bs{\chi}_i \,   \qquad \text{and} \qquad
\bs{\psi}^i \, = \, - \, \bs{D} \bs{X}^i  \ .
 \ee
It gives the gauge fixed BV symplectic structure
\be \label{eq:ExactGaugeBVsymplStdCourant}
\bs{\omega}_{3,\rm{gf}} \, = \, \oint_{T[1]\partial\Sigma_3} \, \dd^2
\hat z \ \bs{\delta} \bs{X}^i \, \bs{\delta} \bs{\chi}_i \ , 
\ee
and reduces the AKSZ action \eqref{eq:AKSZactionStdCourant} without
$H$-flux to zero. With $H$-flux, the AKSZ action leads to a pure
Wess-Zumino coupling
\be \label{eq:pureHfluxterm}
 \mbf \cS{}_{H,\rm{gf}}^{(3)} \, = \, - \, \frac1{3!} \, \int_{T[1]\Sigma_3}\,\dd^3\hat z \ \mbf H_{ijk}\, \mbf D \bs{X}^i\,\mbf D \bs{X}^j \, \mbf D \bs{X}^k \ 
\ee
which is no longer an AKSZ action, as there are no BV gauge degrees of
freedom in the bulk. This is reminescent of the fact that the equation of motion for
$\mbf F{}_i$ also gives the same action \eqref{eq:pureHfluxterm} up to
a sign. If $H_{ijk}=\partial_{[i}B_{jk]}$ is exact, then we
obtain the boundary AKSZ action
\be 
 - \, \frac1{2} \, \oint_{T[1]\partial\Sigma_3}\,\dd^2\hat z \ \mbf B_{ij}\, \mbf D \bs{X}^i\,\mbf D \bs{X}^j \ , 
\ee
which is the quantization of the NS--NS $B$-field coupling. Hence the
exact gauge is nicely applicable for boundary reductions of topological membranes describing flux deformations of string sigma-models.

\medskip

{\underline{\sl Relation to the B-model.} \ }
We will now show that the standard Courant sigma-model is related to the B-model on its boundary via the exact gauge.
Although we have found that the standard Courant sigma-model has a trivial
boundary reduction in the exact gauge, we can obtain a non-trivial
boundary theory if we extend its field content and then set the extra
fields to zero with gauge fixing.

The standard Courant sigma-model without $H$-flux is given by the Hamiltonian $\gamma_0$ in \eqref{eq:HamStdCourant} and the symplectic form $\omega_3$ in \eqref{eq:StdCourantSympl}. The AKSZ action is
\be \label{eq:AKSZactionStdCourant2}
\bs{\cS}_{0}^{(3)} \, = \, \int_{T[1]\Sigma_3} \, \dd^3\hat z \
\big(\bs{F}_i \, \bs{D}\bs{X}^i \, - \, \bs{\chi}_i \, \bs{D}
\bs{\psi}^i \, + \, \bs{F}_i \, \bs{\psi}^i \big)  \ .
\ee
We double its fields with the introduction of degree 0 coordinates $\tx_i$, degree 1 coordinates $\wt{\chi}{}^{\,i}$, $\wt{\psi}{}_i$ and degree 2 coordinates $\wt{F}{}^i$ on the target QP2-manifold. The extra term in the symplectic structure is
\be
 \dd \tx_i \w \dd \wt{F}{}^i \, + \, \dd \wt{\chi}{}^{\,i} \w \dd \wt{\psi}{}_i \ ,
\ee
which together with the extended Hamiltonian
\be
\gamma_0 \, + \, {\widetilde\gamma}_0 \, = \, F_i \, \psi^i \, - \,
\wt{\psi}{}_i \, \wt{F}{}^i
\ee
defines the AKSZ action
\be \label{eq:AKSZactionEnlargedStdCourant}
\bs{\widetilde\cS}_{0}^{\,(3)} \, = \, \int_{T[1]\Sigma_3} \, \dd^3\hat
z \ \big(\bs{F}_i \, \bs{D}\bs{X}^i \, - \, \bs{\chi}_i \, \bs{D}
\bs{\psi}^i \, + \, \bs{F}_i \, \bs{\psi}^i \, - \, \wt{\mbf\psi}{}_i
\, \wt{\bs{F}}{}^i \big) \ ,
\ee
where we did not introduce all the possible kinetic terms. This extended standard Courant sigma-model is comparable to
the membrane sigma-model in~\cite{Ikeda2007}, which was introduced in
order to uplift the AKSZ construction of the B-model in
\eqref{eq:BmodAKSZaction3} to an AKSZ membrane theory with generalized complex structure. Our construction arrives at a different B-model construction and uses less fields, but does not include the generalized complex structure.

The last term in \eqref{eq:AKSZactionEnlargedStdCourant} decouples from the original standard Courant sigma-model.
To see this we can choose a different gauge than that we will choose for the boundary reduction, but we use the same field-antifield decomposition. For example, if we set $\wt{\bs{F}}{}^i=0$ and $\wt{\mbf\chi}{}^i=0$ as a partial gauge fixing, we can trivially integrate out the fields $\mbf\tx{}_i$ and $\wt{\mbf\psi}_i$, which gives the action of the standard Courant sigma-model in \eqref{eq:AKSZactionStdCourant2}.
Alternatively, we can arrive at the same conclusion if we rescale the
fields by a real parameter
$\lambda$ in a way which leaves the symplectic structure invariant:
\be
\wt{\mbf\chi}{}^i \, \longrightarrow \, \lambda \, \wt{\mbf\chi}{}^i
\qquad \text{and} \qquad \wt{\mbf\psi}_i \, \longrightarrow \,
\frac{1}{\lambda} \, \wt{\mbf\psi}_i \ ,
\ee 
which is a duality transformation given by a symplectomorphism at the BV level. Then
we take the $\lambda \rightarrow \infty$ limit: the term
$\wt{\mbf\psi}_i \, \wt{\bs{F}}{}^i$ in the action tends to zero and we
get the standard Courant sigma-model in this way as well. Later on we
will employ a similar rescaling technique.

Now we reduce the extended standard Courant sigma-model to its
boundary with the previously defined exact gauge from \S\ref{sec:gaugefixing}. In this case it means the specific gauge choice
\be 
\wt{\mbf\chi}{}^i \, = \, - \, \bs{D} \bs{X}^i \ , \qquad 
\bs{F}_i \, = \, \bs{D} \wt{\mbf\psi}_i \ , \qquad 
\bs{\chi}_i \, = \, - \, \bs{D} \mbf\tx{}_i \qquad \mbox{and} \qquad
\wt{\bs{F}}{}^i \, = \, \bs{D} \bs{\psi}^i \ .
 \ee
The BV symplectic form becomes
\be
\oint_{T[1]\partial\Sigma_3} \, \dd^2\hat z \ \big(\bs{\delta}\bs{X}^i
\, \bs{\delta} \wt{\mbf\psi}_i \, + \, \bs{\delta} \mbf\tx{}_i \, \bs{\delta} \wt{\mbf\chi}{}^i \big) \ ,
\ee
which is the same BV symplectic form induced by \eqref{eq:BmodSympHofman} and \eqref{eq:BmodSympHofmanEnlarged} with the relabelling $\wt{\mbf\psi}_i \rightarrow \mbf\chi{}_i$. 
Our gauge fixing also reduces the AKSZ action in \eqref{eq:AKSZactionEnlargedStdCourant} to the boundary action
\be
\bs{\widetilde\cS}_{0,\rm{gf}}^{(3)} \, = \,
\oint_{T[1]\partial\Sigma_3} \, \dd^2\hat z \ \big(\wt{\mbf\psi}_i \,
\bs{D}\bs{X}^i \, + \, \mbf\tx{}_{i} \, \bs{D} \wt{\mbf\chi}{}^i \, +
\, \wt{\mbf\psi}_i \, \wt{\mbf\chi}{}^i \big) \, = \,
\bs{\cS}_{\rm{B}1}^{(2)}  \ ,
\ee
which is the same action as that of the B-model in \eqref{eq:BmodAKSZaction1} with the same relabelling $\wt{\mbf\psi}_i \rightarrow \bs{\chi}_i$ as before. 

\subsection{Contravariant Courant sigma-model}
\label{sec:AmodContra}

The contravariant Courant sigma-model was introduced in
\cite{Bessho2015} as the Courant sigma-model corresponding to a
Poisson Courant algebroid. It is defined by the AKSZ action 
\be \label{eq:AKSZContraCourant}
\bs{\cS}_{\pi,R}^{(3)} \, = \, \int_{T[1]\Sigma_3} \, \dd^3 \hat z \
\Big( \bs{F}_i \, \bs{D}\bs{X}^i \, - \, \bs{\chi}_i \, \bs{D}
\bs{\psi}^i \, + \, \bs{\pi}^{ij} \, \bs{F}_i \, \bs{\chi}_j \, - \,
\frac 12 \, \bs{\partial}_i \bs{\pi}^{jk} \, \bs{\psi}^i \,
\bs{\chi}_j \, \bs{\chi}_k \,  + \, \frac{1}{3!} \, \bs{R}^{ijk} \,
\bs{\chi}_i \, \bs{\chi}_j \, \bs{\chi}_k \Big) \ .
\ee
In the absence of $R$-flux the master equation gives the Poisson
condition for the bivector $\pi$, so we can expect that the
contravariant Courant sigma-model is closely related to the Poisson
sigma-model. This relation turns out to be the exact gauge boundary
reduction. We use the same gauge fixing as we used for the standard
Courant sigma-model in \eqref{eq:ExactGaugeStdCourant} which gives the
boundary BV symplectic form
\eqref{eq:ExactGaugeBVsymplStdCourant}. The resulting boundary AKSZ
action is that of the Poisson sigma-model in \eqref{eq:PoissonActAKSZ}:
\be
\bs{\cS}_{\pi,0;\rm{gf}}^{(3)} \, = \, \oint_{T[1]\partial\Sigma_3}  \,
\dd^2 \hat z \ \Big(\bs{\chi}_i \, \bs{D} \bs{X}^i \, + \, \frac 12 \,
\bs{\pi}^{ij} \, \bs{\chi}_i \, \bs{\chi}_j \Big) \, = \, \bs{\cS}_{\pi}^{(2)} \ .
\ee

\medskip

{\underline{\sl Relation to the sigma-model for non-geometric $R$-flux.}
  \ }  
In the degenerate limit where the anchor of the contravariant Courant
sigma-model is set to zero, we show that, in the exact gauge, it
coincides precisely with the membrane sigma-model
of~\cite{Richard2012} which quantizes the nonassociative phase space
and geometry of the $R$-flux
background~\cite{Blumenhagen2010,Lust2010,Blumenhagen2011}. This
clarifies more precisely the geometrical meaning of the model
of~\cite{Richard2012} in terms of a Courant algebroid structure. An
alternative geometric description as a certain reduction of the
standard Courant sigma-model for the target space of double field
theory is discussed in~\cite{Richard2018}, which we study in \S\ref{sec:DFTAlgAndSigmaModel}.
A vanishing anchor map $\rho$ with non-zero $R$-flux means that the
bivector field $\pi$ is identically zero, and the Dorfman bracket is
given solely by the three-vector $R$ in the simple form
\be
[X+\alpha,Y+\beta]_{{\rm D};0,R} \, = \, - \, \iota_\alpha \iota_\beta R \ ,
\ee 
so that the tangent bundle $TM$ decouples completely from this structure.
 
We choose the exact gauge \eqref{eq:ExactGaugeStdCourant}. In this
case our gauge choice is not compatible with boundary conditions,
because the BV master equation forces the flux term to be zero on the
boundary, which means $\bs{\chi}_i=0$ on $T[1]\partial\Sigma_3$ if
$R\neq 0$. This can be circumvented by adding a non-topological
boundary term to the action as in~\cite{Richard2012}. We introduce
this as a strictly classical term after the full gauge fixing, and for brevity
avoid here issues concerning its quantization. 
Hence the AKSZ action \eqref{eq:AKSZContraCourant} with $\pi=0$ reduces to 
\be
\bs{\cS}_{0, R; \rm{gf}}^{(3)} \, = \, \oint_{T[1]\partial\Sigma_3} \,
\dd^2 \hat z \ \bs{\chi}_i \, \bs{D} \bs{X}^i \, + \, \frac{1}{3!} \,
\int_{T[1]\Sigma_3}  \, \dd^3 \hat z \ \bs{R}^{ijk} \, \bs{\chi}_i \,
\bs{\chi}_j \, \bs{\chi}_k \  .
\ee

There is still a gauge degree of freedom on the boundary fields 
remaining, therefore we choose $\chi_i^{(0)}=\chi_i^{(2)}=X^{i\,(1)}=0$. The bulk fields are not antifields in the BV sense, so we cannot set any of them to zero. Instead we eliminate the non-zero parts in the bulk by their equations of motion. The gauge fixed action for constant $R$ in the superfield expansion \eqref{eq:superfieldexp} is
 \be
 \bs{\cS}_{0,R; \rm{gf}}^{(3)} \, = \, \oint_{\partial\Sigma_3} \,
 \dd^2 \sigma \ \chi_i \w \dd X^i \, + \, \frac{1}{3!} \,
 \int_{\Sigma_3} \, \dd^3 \sigma \ \big(  R^{ijk} \, \chi_i \w \chi_j
 \w \chi_k \, - \, R^{ijk} \, f_i \, \chi_j \w \phi_k \, + \, R^{ijk}
 \, f_i \, f_j \, C_k \big)
 \ee
where $\chi^{(1)}_i=\chi_i$ is a degree 0 one-form, $\chi^{(0)}_i=f_i$
is a degree 1 function, $\chi^{(2)}_i=\phi_i$ is a degree $-1$
two-form and $\chi^{(3)}_i=C_i$ is a degree $-2$ three-form. Both
$f_i$ and $\phi_i$ vanish on the boundary, due to the boundary gauge
fixing. The equations of motion of the three non-zero degree fields
sets the last two bulk terms to zero, and they are consistent with
each other. Now we introduce a boundary term given by the inverse of
a target space metric $g^{ij}$ since we need $\chi_i$ to be non-zero on the boundary. 
Finally we arrive at the action containing only degree~0 fields:
\be
\bs{\cS}_{0,R; \rm{gf}}^{(3)} \, = \, \oint_{\partial\Sigma_3} \,
\dd^2 \sigma \ \chi_i \w \dd X^i \, + \, \frac{1}{3!} \,
\int_{\Sigma_3} \, \dd^3 \sigma \ R^{ijk} \, \chi_i \w \chi_j \w
\chi_k \,  + \, \oint_{\partial\Sigma_3} \, \dd^2 \sigma \ g^{ij} \, \chi_i \w * \chi_j \ ,
\ee 
where $\ast$ is the Hodge duality operator corresponding to a chosen
metric on the membrane worldvolume $\Sigma_3$. This is precisely the
string sigma-model derived in~\cite{Richard2012} which quantizes the
non-geometric $R$-flux background. 

\section{DFT membrane sigma-models}
\label{sec:DFTAlgAndSigmaModel}

Double field theory is a manifestly T-duality invariant theory, in
which non-geometric backgrounds can be described naturally. It has its
spacetime doubled, where the original coordinates conjugate to
momentum modes are supplemented by dual coordinates conjugate to
winding modes of closed strings. On the other hand, flux backgrounds
and generalized geometry in closed string theory can be studied within the
context of open topological membrane sigma-models, which uplift string
theory on their boundary to bulk membrane theories. These membrane
sigma-models fit nicely into the framework of the AKSZ construction of
membrane sigma-models: the Courant bracket structure of generalized
diffeomorphisms corresponds to the derived bracket of Courant
sigma-models. The introduction of an analogous correspondence in
double field theory appears in~\cite{Richard2018}, where DFT
algebroids are defined as the appropriate double field theory
analogues of Courant algebroids in generalized geometry; they have
the C-bracket as their bracket operation, and reduce to a Courant
algebroid after imposing the strong section constraints. DFT
algebroids can be implemented within topological membrane
sigma-models, which can be obtained by reducing (or projecting) larger
AKSZ sigma-models. In this section we describe the construction of DFT
membrane sigma-models in AKSZ theory.

\subsection{DFT algebroids} 

The definition of DFT algebroid starts with the large Courant algebroid which is a straightforward doubled version of a general Courant algebroid. We formulate the definition from the graded symplectic geometry viewpoint, since this is explicitly relevant in AKSZ constructions.
For simplicity we consider the case when the large Courant algebroid
is a Courant algebroid corresponding to the QP2-manifold
$T^*[2]T[1]T^* M$, where the doubling of the original base manifold
$M$ appears as the total space of the cotangent bundle $T^* M$. We use
the doubled index $I=1,\ldots,2d$ to label coordinates on the base
space $T^*M$, which can be split into the first $d$ indices
$I=1,\ldots,d$, which are the original indices labelling coordinates
on $M$, and the second $d$ dual indices $I=d+1,\ldots,2d$ labelling
the covectors of $T^*M$; both sets of indices are labeled by
$i=1,\ldots,d$. 

The symplectic form coming from
\eqref{eq:StdCourantSympl} is 
\be \label{eq:LargeCourantsympl} \begin{aligned}
\omega_{\rm{DFT}} & = \, \dd X^I \w \dd F_I \, + \, \dd \chi_I \w \dd \psi^I  \\[4pt]
 & = \,  \dd X^i \w \dd F_i \, + \, \dd \wt{X}{}_i \w \dd \wt{F}{}^i \, + \, \dd \chi_i \w \dd \psi^i \, + \, \dd \wt{\chi}{}^{\,i} \w \dd \wt{\psi}{}_i \ ,
\end{aligned} \ee
where the splitting of a general doubled coordinate $\phi^I$ has been
used: $\phi^I=(\phi^i,\wt{\phi}{}_i)$. As we know from
\eqref{eq:QPmanGeoCorrespondence} a degree one function $A^I \, \chi_I
+ \alpha_I \, \psi^I$ corresponds to a section of $T(T^*M)\oplus
T^*(T^*M)$ symbolically as
\be  
A^i \, \chi_i \, + \,  \wt{A}{}_i \, \wt{\chi}{}^{\,i} \, + \, \alpha_i \, \psi^i \, + \, \wt{\alpha}{}^{\,i} \, \wt{\psi}{}_i \quad \longleftrightarrow \quad A^i
 \, \frac\partial{\partial X^i} \, + \, \wt{A}{}_i \, \frac\partial{\partial \wt{X}{}_i}  \, + \, \alpha_i \, \dd X^i \, + \, \wt{\alpha}{}^{\,i} \, \dd \wt{X}{}_i \ ,
\ee
and the derived brackets \eqref{eq:derivedStrCourantAlg} with a given general Hamiltonian \eqref{eq:HamiltonianCourantAlg}:
\be \begin{aligned} \label{eq:LargeCourantHam}
\gamma_{\rm{DFT}} \, = \, & \rho^I{}_J \, F_I \, \psi^J \, + \, \beta^{IJ} \, F_I \, \chi_J \, + \, \frac{1}{3!} \,
T^{(0)}{}_{IJK} \, \psi^I \, \psi^J \, \psi_K \\
& + \, \frac 12 \, T^{(1)}{}_{IJ}{}^K \, \psi^I \, \psi^J \, \chi_K \, + \, \frac 12 \, T^{(2)}{}_I{}^{JK} \, \psi^I \, \chi_J \, \chi_K \, + \, \frac{1}{3!} \, T^{(3)\,IJK} \, \chi_I \, \chi_J \, \chi_K \ , 
\end{aligned}\ee
define a Courant algebroid on $T(T^*M)\oplus T^*(T^*M)$.

The DFT algebroid is based on the projection to DFT vectors, which
halves the number of degree~1 coordinates. We introduce a new basis
for the subspace of degree 1 fields spanned by $\chi_I$ and $\psi^I$
given as
\be \label{eq:DFTCourantAlgTauDef}
\tau_{\pm}^I \, = \, \frac 12 \, \big(\psi^I \, \pm \, \eta^{IJ} \, \chi_J \big) \ ,
\ee
where $\eta_{IJ}$ is the $O(d,d)$-invariant constant metric
\be\label{eq:Oddmetric}
\eta_{IJ}\, = \, \begin{pmatrix} 0 & \delta^i{}_j \\ \delta_i{}^j & 0 \end{pmatrix} \ .
\ee
The projection $\mathsf{p}_+$ to the subspace $L_+$ spanned by $\tau^I_+$
yields the projection to DFT vectors, which are vectors under
$O(d,d)$. The corresponding sub-bundle of $T(T^*M)\oplus T^*(T^*M)$ is also
denoted~$L_+$. 

For the symplectic structure $\omega_{\rm{DFT}}$ the projection means
\bea
\dd \chi_I \w \dd \psi^I  \, = \, \eta_{IJ} \, \dd \tau_+^I \w \dd
\tau_+^J \, - \, \eta_{IJ} \, \dd \tau_-^I \w \dd \tau_-^J \
\xrightarrow{ \ \mathsf{p}_+ \ } \ \eta_{IJ} \, \dd \tau^I_+ \w \dd \tau^J_+ \ .
\eea
The coordinates $\tau_+^I$ are counted twice in the symplectic
structure compared to the original \eqref{eq:LargeCourantsympl}, so we
halve their contribution in the symplectic structure solely: 
\bea \label{eq:DFTSigmaModSymplRed}
\omega_{\rm{DFT},+} \, = \, \dd X^I \w \dd F_I \, + \, \frac12 \, \eta_{IJ} \, \dd \tau^I_+ \w \dd \tau^J_+ \ .
\eea
For the Liouville potential we take $\vartheta_{\rm{DFT},+} = F_I\,
\dd X^I - \frac12\, \eta_{IJ} \, \tau_+^I\, \dd\tau_+^J$. To specify
its zero locus $\cL_{\rm{DFT},+}$ as a Lagrangian submanifold, we choose a
polarisation which is defined by a projector $\cP$ on $L_+$ of rank
$d$ that is maximally isotropic with respect to the $O(d,d)$ metric
\eqref{eq:Oddmetric}:
\be
\cP^i{}_K\, \eta^{KL}\, \cP^j{}_L = 0 \ .
\ee
It acts
on the basis $\tau_+^I$ to give degree~1 coordinates
\be
\tau^i=\cP^i{}_J\, \tau_+^J
\ee
which span a $d$-dimensional subspace of
$L_+$; then $\cL_{\rm{DFT},+}:=\{F_I=0, \, \tau^i=0\}$. Different polarizations
define different Lagrangian submanifolds which are all related by
$O(d,d)$ transformations: Acting with $\cO\in O(d,d)$ changes the polarization as
\be\label{eq:polarizationchange}
\big(\,{}^\cP_{\widetilde\cP}\,\big) \longmapsto \big(\,{}^{\cP'}_{\widetilde\cP{}'}\,\big) = \big(\,{}^\cP_{\widetilde\cP}\,\big) \, \cO \ ,
\ee
where $\widetilde\cP=1-\cP$ is the complementary projector.

The Hamiltonian is projected to the subspace $L_+$ as
\be
\gamma_{\rm{DFT},+} \, = \,  {(\rho_+)^I}_J \, F_I \, \tau^J_+ \, + \, \frac{1}{3!} \,
(T_+)_{IJK} \, \tau^I_+ \, \tau^J_+ \, \tau^K_+ \ , 
\ee
where the new structure functions are defined by
\be
{(\rho_+)^I}_J \, = \, \rho^I{}_J \, + \, \beta^{IK} \, \eta_{KJ}  \ ,
\ee
and
\be\begin{aligned}
(T_+)_{IJK} \, = \ \, & T^{(0)}{}_{IJK}  \, + \, 3 \,
T^{(1)}{}_{[IJ}{}^{K'} \, \eta_{K]K'} \, + \, 3 \,
T^{(2)}{}_{[I|}{}^{J'K'} \, \eta_{J'|J|} \, \eta_{K'|K]} \\ & + \,
T^{(3)\,I'J'K'} \, \eta_{I'[I|} \, \eta_{J'|J|} \, \eta_{K'|K]} \ .
\end{aligned}\ee
The Hamiltonian $\gamma_{\rm{DFT},+}$ defined by these functions does
not necessarily satisfy the master equation, despite the fact that the original Hamiltonian $\gamma_{\rm{DFT}}$ of the large Courant algebroid defined by $\rho^I{}_J$, $\beta^{IJ}$, $T^{(0)}{}_{IJK}$, $T^{(1)}{}_{IJ}{}^K$, $T^{(2)}{}_{I}{}^{JK}$ and $T^{(3)\,IJK}$  does.

The C-bracket is defined on DFT vectors of $L_+$, which correspond to
the degree~1 functions $A$ in the subspace spanned by $e_I=\eta_{IJ} \, \tau_+^J$:
\be
A \, = \, A^I \, e_I \, = \, \frac12 \, A^I \, \big(\chi_I + \eta_{IJ} \, \psi^J\big) \ .
\ee
It can be obtained from derived brackets of the QP2-manifold together
with the symmetric pairing and anchor map as
 \be \begin{aligned}
  \llbracket A_1,A_2 \rrbracket_{L_+} &=  - \, \frac 12 \, \big(\{\{ A_1 , \gamma_{\rm{DFT},+} \} , A_2 \} \, - \, \{\{ A_2 , \gamma_{\rm{DFT},+} \} , A_1 \} \big) \ , \\[4pt]
 \langle A_1, A_2\rangle_{L_+}  &= \{ A_1,A_2 \} \ , \\[4pt]
  \rho_+(A) &=  \{ A , \{ \gamma_{\rm{DFT},+} , \, \cdot \, \}\} \ ,
\end{aligned}\ee
for DFT vectors $A$, $A_1$ and $A_2$. Since the master equation does
not hold for $\gamma_{\rm{DFT},+}$, the algebraic structure is
not that of a Courant algebroid, but called a DFT algebroid
in~\cite{Richard2018}:
A \emph{DFT algebroid} on $T^*M$ is a vector bundle $L_+$ of rank $2d$
over $T^* M$ equiped with a non-degenerate symmetric form $\langle\,
\cdot \, , \, \cdot \, \rangle_{L_+} $ on its fibres, an anchor map $\rho_+ :
L_+ \rightarrow T(T^* M)$, and a skew-symmetric bracket of sections
$\llbracket \, \cdot \, , \, \cdot \, \rrbracket_{L_+}$, called the
C-bracket, which together satisfy
\be \begin{aligned}
\langle \mathcal{D}_+ f , \mathcal{D}_+ g \rangle_{L_+} &= \, \frac 14
\, \langle \dd f , \dd g \rangle_{L_+} \ , \\[4pt]
\llbracket A , f \, B \rrbracket_{L_+} &= \, f \, \llbracket A, B
\rrbracket_{L_+} \, + \, \big(\rho_+(A)f\big) \, B \, - \, \langle
A,B\rangle_{L_+} \, \mathcal{D}_+ f  \ , \\[4pt]
\rho_+ (C) \langle A , B \rangle_{L_+} &= \, \big\langle \llbracket C,A \rrbracket_{L_+} \, + \, \mathcal{D}_+
\langle C,A \rangle_{L_+} \,,\, B \big\rangle_{L_+} \, + \, \big\langle A \,,\, \llbracket C,B \rrbracket_{L_+} \, + \, \mathcal{D}_+ \langle C,B \rangle_{L_+} \big\rangle_{L_+} \ , 
\end{aligned}\ee
for all sections $A,B,C\in C^\infty(T^*M,L_+)$ and functions $f,g\in
C^\infty (T^* M)$, where $\mathcal{D}_+ : C^\infty(T^* M) \rightarrow
C^\infty(T^*M,L_+) $ is the derivative defined through $\langle
\mathcal{D}_+ f , A \rangle_{L_+} = \frac 12 \, \rho_+ (A) f$.

\subsection{AKSZ construction of DFT membrane sigma-models}

The large Courant sigma-model constructed by AKSZ theory corresponds to the
large Courant algebroid in the spirit of
\S\ref{sec:CourantSigmaMod}. Then we execute the projection by
$\mathsf{p}_+$ on the level of AKSZ fields, which means selecting a
special submanifold (by projection to DFT vectors) containing half of the
ghost number~1 fields. This method is quite natural, because fields with identical properties appear twice, and we keep only one field of each identical pair. 
Note that there are infinitely many possibilities to perform the
reduction on ghost number~1 fields, but only this projection to DFT
vectors gives the right C-bracket structure of double field
theory. One can think of the other reductions as a class of duality
transformations, which leads out of the realm of the original physical
double field theory. 

The AKSZ action of the large Courant sigma-model corresponding to \eqref{eq:LargeCourantHam} is defined by
\be \begin{aligned}
\bs{\cS}_{\rm{DFT}}^{(3)} \, = \, \int_{T[1]\Sigma_3} \, \dd^3 \hat z
\ \Big( & \bs{F}_I \, \bs{D} \bs{X}^I \, - \, \bs{\chi}_I \, \mbf D \bs{\psi}^I \, + \, \bs{\rho}^I{}_J \, \bs{F}_I \, \bs{\psi}^J \, + \, \bs{\beta}^{IJ} \, \bs{F}_I \, \bs{\chi}_J \, + \, \frac{1}{3!} \,
\bs{T}^{(0)}{}_{IJK} \, \bs{\psi}^I \, \bs{\psi}^J \, \bs{\psi}_K \\
& + \, \frac 12 \, \bs{T}^{(1)}{}_{IJ}{}^K \, \bs{\psi}^I \, \bs{\psi}^J \, \bs{\chi}_K \, + \, \frac 12 \, \bs{T}^{(2)}{}_I{}^{JK} \, \bs{\psi}^I \, \bs{\chi}_J \, \bs{\chi}_K \, + \, \frac{1}{3!} \, \bs{T}^{(3)\,IJK} \, \bs{\chi}_I \, \bs{\chi}_J \, \bs{\chi}_K \Big) \ .
\end{aligned}\ee
The BV symplectic structure coming from \eqref{eq:LargeCourantsympl} is given by
\be \label{eq:LargeCourantBVsympl} \begin{aligned}
\bs{\omega}_{\rm{DFT}} & = \, \int_{T[1]\Sigma_3} \, \dd^3 \hat z \ \big(  \mbf\delta \bs{X}^I \, \mbf\delta \bs{F}_I \, + \, \mbf\delta \bs{\chi}_I \, \mbf\delta \bs{\psi}^I \big) \\[4pt]
 & = \, \int_{T[1]\Sigma_3} \, \dd^3 \hat z \ \big(  \mbf\delta \bs{X}^I \, \mbf\delta \bs{F}_I \, + \, \eta_{IJ} \, \mbf\delta \bs{\tau}^I_+ \, \mbf\delta \bs{\tau}^J_+ \, - \, \eta_{IJ} \, \mbf\delta \bs{\tau}^I_- \, \mbf\delta \bs{\tau}^J_- \big) \ ,
\end{aligned} \ee
where we performed a duality transformation originating from \eqref{eq:DFTCourantAlgTauDef}:
\be
\bs{\tau}_{\pm}^I \, = \, \frac 12 \, \big(\bs{\psi}_I \, \pm \, \eta_{IJ} \, \bs{\chi}^J \big) \ .
\ee

Now we restrict the superfields to the submanifold
$\bs{\tau}_{-}^I=0$. This is not a partial gauge fixing, since both
fields and antifields are set to zero. The corresponding action is
given by
\be 
\bs{\cS}_{\rm{DFT},+}^{(3)} \, = \,  \int_{T[1]\Sigma_3} \, \dd^3 \hat
z \ \Big( \bs{F}_I \, \bs{D} \bs{X}^I \, - \, \eta_{IJ} \,
\bs{\tau}^I_+ \, \mbf D \bs{\tau}^J_+ \, + \, (\bs{\rho}_+)^I{}_J \, \bs{F}_I \, \bs{\tau}_+^J \, + \, \frac{1}{3!} \, (\bs{T}_+)_{IJK} \, \bs{\tau}^I_+ \, \bs{\tau}^J_+ \, \bs{\tau}^K_+ \Big) \ .
 \ee
The reason we do not define this action directly from a DFT algebroid
is that the action $\bs{\cS}_{\rm{DFT},+}^{(3)}$ does not satisfy the
BV master equation, so it cannot be constructed by AKSZ theory, nor
does it define a BV quantized sigma-model. In order for it to define a
BV quantized theory or an AKSZ theory we have to impose additional
conditions on the structure functions $(\rho_+)^I{}_J$ and
$(T_+)_{IJK}$ coming from the BV master equation for the reduced
action. We will use this method below in our study of the topological
A- and B-models within the framework of double field theory. 

\section{Double field theory for the A- and B-models}
\label{sec:UnifyABDFT}

There have been several attempts to find an AKSZ sigma-model for both
the topological A- and B-models, which unifies them within the
framework of generalized geometry. These were based on AKSZ
constructions for generalized complex geometry, which describes the
K{\"a}hler structure of A-model and the complex structure of the
B-model within one generalized complex structure. In the remainder of
this paper we continue their
study within the context of double field theory.

\subsection{Doubled Poisson sigma-model}
\label{sec:doubledPoisson}

We shall start by proposing an AKSZ Poisson sigma-model with doubled target space coordinates, which gives the A- and B-models separately. 
Let us consider an AKSZ Poisson sigma-model with target QP1-manifold
$T^*[1]T^*M$ with degree~0 and degree~1 coordinates
\be
X^I \, = \, \begin{pmatrix} X^i \\ \tx_i \end{pmatrix} \qquad \mbox{and} \qquad \chi_I \, = \, \begin{pmatrix} \chi_i \\ \wt{\chi}{}^{\,i} \end{pmatrix} 
\ee
respectively. The doubled Poisson structure is denoted by
$\Omega^{IJ}$, and it depends on both degree~0 coordinates $X^i$ and
$\wt{X}_i$. The symplectic form and AKSZ action are as introduced in \S\ref{sec:AKSZPoisson}:
\be
\omega_{\rm{D}2} \, = \, \dd \chi_I \w \dd X^I \ ,
\ee
and
\be
\bs{\cS}_{\Omega}^{(2)} \, = \, \int_{T[1]\Sigma_2} \, \dd^2\hat z \
\Big(\bs{\chi}_I \, \bs{D} \bs{X}^I \, + \, \frac 12 \,
\bs{\Omega}^{IJ} \, \bs{\chi}_I \, \bs{\chi}_J \Big) \ .
\ee
The master equation imposes the Poisson condition for $\Omega^{IJ}$:
\be\label{eq:DPoissonConstr}
\Omega^{[I|L} \, \partial_L \Omega^{JK]} \, = \, 0 \ ,
\ee
where the doubled derivative $\partial_I$ is defined by
\be
\partial_I \, = \, \begin{pmatrix} \Large \sfrac{\partial}{\partial X^i} \normalsize \vspace{1mm}  \\ \Large \sfrac{\partial}{\partial \tx{}_i} \normalsize \end{pmatrix} \ . 
\ee
We will now show that particular choices of $\Omega^{IJ}$ give the A- or B-models.

\medskip

{\underline{\sl A-model.} \ } 
The A-model is obtained by using the doubled Poisson structure
\be \label{eq:DPoissonA}
\Omega^{IJ}_{\rm{A}} \, = \, \begin{pmatrix} \pi^{ij} & 0 \\ 0 & 0 \end{pmatrix} \ ,
\ee
where the bivector $\pi^{ij}$ only depends on $X^i$. The AKSZ action $\bs{\cS}_{\Omega_{\rm{A}}}^{(2)}$ defined by $\Omega^{IJ}_{\rm{A}}$ is given by
\be
 \bs{\cS}_{\Omega_{\rm{A}}}^{(2)} \, = \, \int_{T[1]\Sigma_2} \, \dd^2
 \hat z \ \Big(\bs{\chi}_i \, \bs{D} \bs{X}^i \, + \,
 \wt{\bs{\chi}}{}^i \, \bs{D} \wt{\bs{X}}_i \, + \, \frac 12 \,
 \bs{\pi}^{ij} \, \bs{\chi}_i \, \bs{\chi}_j \Big) \ ,
\ee
where the additional term $\wt{\bs{\chi}}{}^i \, \bs{D} \mbf \tx{}_i$
can be removed with a partial gauge fixing. Thus it yields the
original Poisson sigma-model, which is the AKSZ construction of the
A-model \eqref{eq:AmodPoissonAKSZaction}. Then the constraint
\eqref{eq:DPoissonConstr} reduces to the original constraint that
$\pi$ defines a Poisson structure on $M$.

\medskip

{\underline{\sl B-model.} \ } 
The B-model cannot be obtained simultaneously with the A-model, as it
arises from a different doubled Poisson structure. The AKSZ construction given by a constant complex structure in \eqref{eq:BmodAKSZaction3} can be obtained using 
\be \label{eq:DPoissonBconstcomplex}
\Omega^{IJ}_{\rm{B}} \, = \, \begin{pmatrix} 0 & {J^i}_j \\ -{J^j}_i & 0 \end{pmatrix} \ .
\ee
It gives the AKSZ construction of the B-model after the sign flip $\mbf \tx{}_i \rightarrow -\mbf \tx{}_i$.
The AKSZ construction of the B-model in \eqref{eq:BmodAKSZaction1} can be derived directly with the choice $J^i{}_j=\ii\delta^i_j$ from the doubled Poisson sigma-model using
\be \label{eq:DPoissonBsimple}
\Omega^{\prime\,IJ}_{\rm{B}} \, = \, \begin{pmatrix} 0 & \delta^i{}_j \\ -\delta^j{}_i & 0 \end{pmatrix} \ .
\ee

The AKSZ formulation of the B-model for an arbitrary complex structure
${J^i}_j$, which only depends on $X^i$, is given in
\eqref{eq:ComplexStrAKSZaction}. Our doubled Poisson sigma-model also
includes this construction and it is given by choosing 
\be \label{eq:DPoissonCompl}
\Omega^{IJ}_{J} \, = \, \begin{pmatrix} 0 & {J^i}_j \\ -{J^j}_i &
  -2\,\partial_{[i} {J^k}_{j]} \, \tx{}_k \end{pmatrix} 
\ee
after the sign flip $\mbf \tx{}_i \rightarrow - \mbf \tx{}_i$. The constraint \eqref{eq:DPoissonConstr} gives the same constraint as in the original construction, which is the integrability condition \eqref{eq:ComplStrComp}.\footnote{Since the
  complex structure appears as a Poisson structure
  on doubled space, it can be quantized using
  the Cattaneo-Felder approach~\cite{CattFeld1999}. In particular, the
  case of a constant complex structure can be quantized in closed form
  analogously to the Moyal star-product.}

\medskip

{\underline{\sl General doubled Poisson structure.} \ }
We write a general doubled Poisson structure in the block matrix form 
\be \label{eq:DPoissonGen}
\Omega^{IJ}_{\rm{G}} \, = \, \begin{pmatrix} \mathsf{P}^{ij} & {\mathsf{J}^i}_j \\ -{\mathsf{J}^j}_i & \mathsf{Q}_{ij} \end{pmatrix} \ ,
\ee
where the blocks $\mathsf{P}^{ij}$, ${\mathsf{J}^i}_j$ and
$\mathsf{Q}_{ij}$ are constrained by \eqref{eq:DPoissonConstr}, and
they depend on both $X^i$ and $\tx{}_i$.
The corresponding AKSZ action is
\be \label{DPoissonGenAction2D}
 \bs{\cS}_{\Omega_{\rm{G}}}^{(2)} \, = \, \int_{T[1]\Sigma_2} \, \dd^2
 \hat z \ \Big(\bs{\chi}_i \, \bs{D} \bs{X}^i \, + \,
 \wt{\bs{\chi}}{}^i \, \bs{D} \wt{\bs{X}}_i \, + \, \frac 12 \,
 \bs{\mathsf{P}}^{ij} \, \bs{\chi}_i \, \bs{\chi}_j \, + \,
 {\bs{\mathsf{J}}^i}_j \, \bs{\chi}_i \, \wt{\bs{\chi}}{}^j  \, + \,
 \frac 12 \, \bs{\mathsf{Q}}_{ij} \, \wt{\bs{\chi}}{}^i \, \wt{\bs{\chi}}{}^j \Big) \ .
\ee
It is tempting to try to relate $\Omega^{IJ}_{\rm{G}}$ to the general
complex structure ${\mathbb{J}^I}_J$ in \eqref{eq:GenComplexStrMatrix}, but the identities \eqref{eq:GenComplexStrId} are not equivalent to \eqref{eq:DPoissonConstr}. We will return to this problem later. 
It is also interesting to note that the action
$\bs{\cS}_{\Omega_{\rm{G}}}^{(2)}$ gives the Zucchini model
\eqref{eq:ZucchiniAction} if we replace $\wt{\bs{\chi}}{}^i$ with
$\mbf D \bs{X}^i$, and nothing depends on $\tx_i$. But they are
different BV theories with different constraints on their block structures.

\subsection{Large contravariant Courant sigma-model}

We have seen that the A- and B-models on the AKSZ level appear to be
two different particular cases of the same two-dimensional AKSZ theory
on a doubled target space. These Poisson sigma-models can be uplifted
to the membrane level as a contravariant Courant sigma-model from
\S\ref{sec:AmodContra}, which gives them back on the boundary in the
exact gauge. The novelty of the membrane description is that one can
introduce flux terms in the bulk. We shall now study the doubled
contravariant Courant sigma-model with doubled Poisson structures
introduced in \S\ref{sec:doubledPoisson}. These AKSZ constructions will
play a similar role in the study of the A- and B-models within double
field theory as the large Courant sigma-model in \S\ref{sec:DFTAlgAndSigmaModel}.

The BV symplectic form of the contravariant Courant sigma-model in doubled space with QP2-manifold $T^*[2]T[1]T^*M$ is given by \eqref{eq:LargeCourantBVsympl}.
The AKSZ action comes from \eqref{eq:AKSZContraCourant} and is given by
\be  \label{eq:DoubleContraCourantSigmamodelGenAction} \begin{aligned}
\bs{\cS}_{\Omega,\cR}^{(3)} \, = \, \int_{T[1]\Sigma_3} \, \dd^3 \hat
z \ \Big( & \bs{F}_I \, \bs{D}\bs{X}^I \, - \, \bs{\chi}_I \, \bs{D}
{\bs{\psi}}^I \, + \, \bs{\Omega}^{IJ} \, \bs{F}_I \, \bs{\chi}_J \\
& - \, \frac 12 \, \bs{\partial}_I \bs{\Omega}^{JK} \, \bs{\psi}^I \,
\bs{\chi}_J \, \bs{\chi}_K \, + \, \frac{1}{3!} \, \bs{\cR}^{IJK} \,
\bs{\chi}_I \, \bs{\chi}_J \, \bs{\chi}_K \Big) \ ,
\end{aligned}\ee
with the definition of a general three-vector flux $\cR^{IJK}$ on $T^* M$, which is allowed in the contravariant Courant sigma-model.

\medskip

{\underline{\sl A-model.} \ } 
The doubled Poisson structure $\Omega^{IJ}_{\rm{A}}$ in
\eqref{eq:DPoissonA} gives the original contravariant Courant
sigma-model action \eqref{eq:AKSZContraCourant}  on $M$ after the
gauge fixing $\wt{\bs{F}}{}^i = 0 $ and $\wt{\bs{\chi}}{}^i= 0$, which
leaves only the $R$-flux $\cR^{ijk}=R^{ijk}$. One needs to assume that
$R^{ijk}$ only depends on $X^i$ in order to reduce the action purely
to $M$. We have already seen that the contravariant Courant
sigma-model in the exact gauge further reduces to the Poisson
sigma-model formulation of the A-model in \S\ref{sec:AmodContra}.

\medskip

{\underline{\sl B-model.} \ } 
The doubled Poisson structure $\Omega^{\prime\,IJ}_{\rm{B}}$ defined
in \eqref{eq:DPoissonBsimple} with vanishing flux $\cR=0$ gives the
standard Courant sigma-model and a `dual' standard Courant sigma-model
with action
\be
\bs{\cS}_{\Omega'_{\rm{B}},0}^{(3)} \, = \, \int_{T[1]\Sigma_3} \, \dd^3
\hat z \ \big( \bs{F}_i \, \bs{D} \bs{X}^i \, + \, \wt{\bs{F}}{}^i \,
\bs{D} \mbf \tx{}_i \, - \, \bs{\chi}_i \, \bs{D} \bs{\psi}^i \, - \,
\wt{\bs{\chi}}{}^i \, \bs{D} \wt{\bs{\psi}}{}_i \, + \, \bs{F}_i \,
\wt{\bs{\chi}}{}^i \, - \, \wt{\bs{F}}{}^i \, \bs{\chi}_i \big) \ .
\ee
It can be seen that the standard and the dual standard Courant sigma-models are
decoupled from each other (in both the action and the symplectic
form), so they can be separately gauge fixed. For example, the gauge
$\wt{\bs{F}}{}^i = 0 $ and $\bs{\psi}^i = 0 $ yields the standard
Courant sigma-model, which is related to the B-model as described in \S\ref{sec:StdCourantAlgAKSZ}.
The exact gauge defined in \S\ref{sec:gaugefixing} reads here as
\be \begin{aligned} \label{eq:exactgaugeDBmodsimple}
\bs{F}_i \, = \, \mbf D \bs{\chi}_i \ , \qquad \bs{\psi}^i \, = \, - \, \mbf D \bs{X}^i \ , \qquad
\wt{\bs{F}}{}^i  \, = \, \mbf D \wt{\bs{\chi}}{}^i \qquad \mbox{and}
\qquad \wt{\bs{\psi}}{}_i  \, = \, - \, \mbf D \mbf \tx{}_i \ ,
\end{aligned}\ee
and it gives the action of the B-model \eqref{eq:BmodAKSZaction1} on the boundary:
\be
\bs{\cS}_{\Omega'_{\rm{B}},0;\rm{gf}}^{(3)} \, = \,
\oint_{T[1]\partial\Sigma_3} \, \dd^2 \hat z \ \big(\bs{\chi}_i \,
\bs{D}\bs{X}^i \, + \, \mbf \tx{}_{i} \, \bs{D} \wt{\bs{\chi}}{}^{i}
\, + \, \bs{\chi}_{i} \, \wt{\bs{\chi}}{}^{i} \big) \, = \, \bs{\cS}_{\rm{B}1}^{(2)} \ .
\ee

The B-model construction with general constant complex structure is similar: the doubled Poisson structure $\Omega^{IJ}_{\rm{B}}$ in \eqref{eq:DPoissonBconstcomplex} leads to the AKSZ action
\be
\bs{\cS}_{\Omega_{\rm{B}},0}^{(3)} \, = \, \int_{T[1]\Sigma_3} \, \dd^3
\hat z \ \big( \bs{F}_i \, \bs{D} \bs{X}^i \, + \, \wt{\bs{F}}{}^i \,
\bs{D} \mbf \tx{}_i \, - \, \bs{\chi}_i \, \bs{D} \bs{\psi}^i \, - \,
\wt{\bs{\chi}}{}^i \, \bs{D} \wt{\bs{\psi}}{}_i \, + \, J^i{}_j \,
\bs{F}_i \, \wt{\bs{\chi}}{}^j \, - \, J^j{}_i \, \wt{\bs{F}}{}^i \, \bs{\chi}_j \big) \ .
\ee
The exact gauge \eqref{eq:exactgaugeDBmodsimple} gives the B-model construction \eqref{eq:BmodAKSZaction3} on the boundary:
\be
\bs{\cS}_{\Omega_{\rm{B}},0;\rm{gf}}^{(3)}\, = \,
\oint_{T[1]\partial\Sigma_2} \, \dd^2 \hat z \ \big(\bs{\chi}_i \,
\bs{D}\bs{X}^i \, + \, \mbf \tx{}_{i} \, \bs{D} \wt{\bs{\chi}}{}^{i}
\, + \, {J^i}_j \, \bs{\chi}_{i} \, \wt{\bs{\chi}}{}^{j} \big) \, = \, \bs{\cS}_{\rm{B}2}^{(2)} \ .
\ee

Finally, we consider the choice of doubled Poisson structure
$\Omega_{J}$ from \eqref{eq:DPoissonCompl} which is associated to a non-constant complex structure.
The corresponding AKSZ action is given by
\be \begin{aligned} 
\bs{\cS}_{\Omega_{J},0}^{(3)} \, = \, \int_{T[1]\Sigma_3} \, \dd^3 \hat
z \ \big( & \bs{F}_i \, \bs{D} \bs{X}^i \, + \, \wt{\bs{F}}{}^i \,
\bs{D} \mbf \tx{}_i \, - \, \bs{\chi}_i \, \bs{D} \bs{\psi}^i \, - \,
\wt{\bs{\chi}}{}^i \, \bs{D} \wt{\bs{\psi}}{}_i \, + \, {\bs{J}^i}_j
\, \bs{F}_i \, \wt{\bs{\chi}}{}^j \, - \, {\bs{J}^j}_i \,
\wt{\bs{F}}{}^i \, \bs{\chi}_j \\
& - \, 2 \, \bs{\partial}_{[ i } {\bs{J}^k}_{ j ]} \, \mbf\tx{}_k \,
\wt{\bs{F}}{}^i \, \wt{\bs{\chi}}{}^j \, - \, \bs{\partial}_i
{\bs{J}^j}_k \, \bs{\psi}^i \, \bs{\chi}_j \, \wt{\bs{\chi}}{}^k \, +
\, \bs{\partial}_{[i} {\bs{J}^k}_{j]} \, \wt{\bs{\psi}}{}_k \,
\wt{\bs{\chi}}{}^i \, \wt{\bs{\chi}}{}^j \\
& + \, \bs{\partial}_i \bs{\partial}_j {\bs{J}^l}_k \, \mbf\tx{}_l \,
\bs{\psi}^i \, \wt{\bs{\chi}}{}^j \, \wt{\bs{\chi}}{}^k \big) \ .
\end{aligned}\ee
The master equation does not give the integrability condition
\eqref{eq:ComplStrComp} for ${J^i}_j$ this time. We will use the DFT projection later to obtain the right Courant algebroid whose relations give the integrability condition.
In the exact gauge \eqref{eq:exactgaugeDBmodsimple}, the action $\bs{\cS}_{\Omega_{J},0}^{(3)}$ reduces to the boundary action for the non-constant complex structure defined in \eqref{eq:ComplexStrAKSZaction} after a sign flip:
\be
\bs{\cS}_{\Omega_{J},0;\rm{gf}}^{(3)} \, = \,
\oint_{T[1]\partial\Sigma_3} \, \dd^2 \hat z \ \big(\bs{\chi}_i \,
\bs{D}\bs{X}^i \, + \, \mbf \tx{}_{i} \, \bs{D} \wt{\bs{\chi}}{}^{i}
\, + \, {\mbf J^i}_j \, \bs{\chi}_{i} \, \wt{\bs{\chi}}{}^j \, -
\,\bs{\partial}_j \bs{J}{}^i{}_k \, \mbf \tx{}_i \, \wt{\bs{\chi}}{}^j
\, \wt{\bs{\chi}}{}^k \big) \, = \, \bs{\cS}_{J}^{(2)} \ .
\ee

\medskip

{\underline{\sl General doubled Poisson structure.} \ } 
The general AKSZ action
\eqref{eq:DoubleContraCourantSigmamodelGenAction} can be expanded in
block form using the general doubled Poisson structure
$\Omega^{IJ}_{\rm{G}}$ from \eqref{eq:DPoissonGen} as
\be \begin{aligned} \label{eq:LargeCourantSigmaModGen}
\bs{\cS}_{\Omega_{\rm{G}},0}^{(3)} \, =  \, \int_{T[1]\Sigma_3} \, &
\dd^3 \hat z \ \Big( \bs{F}_i \, \bs{D} \bs{X}^i \, + \,
\wt{\bs{F}}{}^i \, \bs{D} \mbf \tx{}_i \, - \, \bs{\chi}_i \, \bs{D}
\bs{\psi}^i \, - \, \wt{\bs{\chi}}{}^i \, \bs{D} \wt{\bs{\psi}}{}_i \,
+ \, \bs{\mathsf{P}}^{ij} \, \bs{F}_i \, \bs{\chi}_j \, + \,
\bs{\mathsf{Q}}_{ij} \, \wt{\bs{F}}{}^i \, \wt{\bs{\chi}}{}^j \\
& + \, {\bs{\mathsf{J}}^i}_j \, \bs{F}_i \, \wt{\bs{\chi}}{}^j \, - \,
{\bs{\mathsf{J}}^j}_i \, \wt{\bs{F}}{}^i \, \bs{\chi}_j \, - \, \frac 12
\, \bs{\partial}_i \bs{\mathsf{P}}^{jk} \, \bs{\psi}^i \,
\bs{\chi}_j \, \bs{\chi}_k \, - \, \frac 12 \, \bs{\partial}_i
\bs{\mathsf{Q}}_{jk} \, \bs{\psi}^i \, \wt{\bs{\chi}}{}^j \,
\wt{\bs{\chi}}{}^k \\ & - \, \bs{\partial}_i {\bs{\mathsf{J}}^j}_k \,
\bs{\psi}^i \, \bs{\chi}_j \, \wt{\bs{\chi}}{}^k
\, - \, \frac 12 \, \wt{\bs{\partial}}{}^i\bs{\mathsf{P}}^{jk} \,
\wt{\bs{\psi}}{}_i \, \bs{\chi}_j \, \bs{\chi}_k \, - \, \frac 12 \,
\wt{\bs{\partial}}{}^i\bs{\mathsf{Q}}_{jk} \, \wt{\bs{\psi}}{}_i \,
\wt{\bs{\chi}}{}^j \, \wt{\bs{\chi}}{}^k \, - \,
\wt{\bs{\partial}}{}^i\bs{\mathsf{J}}^j{}_k \, \wt{\bs{\psi}}{}_i \,
\bs{\chi}_j \, \wt{\bs{\chi}}{}^k \Big) \ ,
\end{aligned}\ee
where the dual derivative is defined by $\wt{\partial}{}^i={\partial}/{\partial \tx{}_i}$. As expected it reduces in the exact gauge \eqref{eq:exactgaugeDBmodsimple} on the boundary to the action $\bs{\cS}_{\Omega_{\rm{G}}}^{(2)}$ given by \eqref{DPoissonGenAction2D}. 

\medskip

{\underline{\sl Fluxes in the A- and B-models.} \ } 
Introducing $\cR$-flux in \eqref{eq:DoubleContraCourantSigmamodelGenAction} gives four different terms
\be \label{eq:TopStrFluxes}
\frac{1}{3!} \, \bs{\cR}^{IJK} \, \bs{\chi}_I \, \bs{\chi}_J \,
\bs{\chi}_K \, = \, \frac{1}{3!} \, \bs{R}^{ijk} \, \bs{\chi}_i \,
\bs{\chi}_j \, \bs{\chi}_k \, + \, \frac{1}{2} \, {\bs{Q}_i}^{jk} \,
\wt{\bs{\chi}}{}^i \, \bs{\chi}_j \, \bs{\chi}_k \, + \, \frac{1}{2}
\, {\bs{F}_{ij}}^{k} \, \wt{\bs{\chi}}{}^i \, \wt{\bs{\chi}}{}^j \,
\bs{\chi}_k \, + \, \frac{1}{3!} \, \bs{H}_{ijk} \, \wt{\bs{\chi}}{}^i
\, \wt{\bs{\chi}}{}^j \, \wt{\bs{\chi}}{}^k \ .
\ee
Both geometric and non-geometric fluxes can appear in the membrane
formulations of the A- and B-models, but the gauge fixings leave only $\bs{R}^{ijk}$ in the case of the original contravariant Courant sigma-model and $\bs{H}_{ijk}$ in the case of the standard Courant sigma-model.
One of the main features of our new construction for the A- and B-models
is that it allows for the introduction of four different fluxes. 
The compatibility condition $[ \Omega , \cR ]_{\rm{S}}=0$ between the
fluxes and the doubled Poisson bivector $\Omega$ can be derived from
the master equation \eqref{eq:ContraCourantMasterConstr}. The same
fluxes \eqref{eq:TopStrFluxes} can be defined in the AKSZ theories
\eqref{eq:LargeCourantSigmaModGen} as well.

\subsection{Courant sigma-model for generalized complex geometry}
\label{sec:DFTtoGenComplStrAB}

So far we have introduced a contravariant Courant sigma-model on
doubled space, which reduces to the topological A- and B-models in the
exact gauge. We shall now treat it as a large Courant sigma-model and
use the projection to DFT vectors from
\S\ref{sec:DFTAlgAndSigmaModel}, which halves the number of degree~1 coordinates. 
Explicitly, the degree~1 coordinates $\chi_I$ and $\psi^I$ are
transformed to $\tau_{\pm}^I$ in \eqref{eq:DFTCourantAlgTauDef}, which in components can be written as
\be 
\tau_{\pm}^I \, = \, \frac 12 \, \begin{pmatrix} \psi^i \pm \wt{\chi}^{\,i}  \\ \wt{\psi}_i \pm \chi_i  \end{pmatrix} \ .
\ee
The coordinates $\tau_{-}^I$ are projected out by $\frac12\,\mathsf{p}_+$ in the same way they were in \eqref{eq:DFTSigmaModSymplRed}, hence in Darboux coordinates
\be
\tau_{+}^I \, = \, \begin{pmatrix} q^i \\ p_i \end{pmatrix}
\ee
the symplectic structure $\omega_{\rm{DFT}}$ from \eqref{eq:LargeCourantsympl} becomes 
\be
\omega_{\rm{DFT}} \ \xrightarrow{ \ \frac 12 \, \mathsf{p}_+ \ } \ \omega_{\rm{DFT},+} \, = \, \dd X^i \w \dd F_i \, + \, \dd \wt{X}_i \w \dd \wt{F}^i \, + \, \dd q^i \w \dd p_i \ .
\ee
In the Hamiltonian \eqref{eq:LargeCourantHam} we substitute
\be \begin{aligned}
\chi_i \, \longrightarrow \, p_i \ , \qquad  \psi^i \, \longrightarrow \, q^i \ ,  \qquad
\wt{\chi}{}^{\,i} \, \longrightarrow \ q^i \qquad \mbox{and} \qquad \wt{\psi}_i \, \longrightarrow \, p_i \ ,
\end{aligned}\ee
which together with the symplectic form $\omega_{\rm{DFT},+}$ reduces
the AKSZ action $\bs{\cS}_{\Omega_{\rm{G}},0}^{(3)}$ from \eqref{eq:LargeCourantSigmaModGen} to the action
\be \begin{aligned} \label{eq:DoubledDFTSigmaModActionAB}
\bs{\cS}_{\Omega_{\rm{G}},+}^{(3)} \, =  \, \int_{T[1]\Sigma_3} & \,
\dd^3 \hat z \ \Big( \bs{F}_i \, \bs{D} \bs{X}^i \, + \,
\wt{\bs{F}}{}^i \, \bs{D} \mbf \tx{}_i \, - \, \bs{p}_i \, \bs{D}
\bs{q}^i  \, + \, \bs{\mathsf{P}}^{ij} \, \bs{F}_i \, \bs{p}_j \, + \,
\bs{\mathsf{Q}}_{ij} \, \wt{\bs{F}}{}^i \, \bs{q}^j \\
& + \, {\bs{\mathsf{J}}^i}_j \, \bs{F}_i \, \bs{q}^j \, - \,
{\bs{\mathsf{J}}^j}_i \, \wt{\bs{F}}{}^i \, \bs{p}_j \, - \, \frac 12
\, \bs{\partial}_i \bs{\mathsf{P}}^{jk} \, \bs{q}^i \, \bs{p}_j \,
\bs{p}_k \, - \, \frac 12 \, \bs{\partial}_i \bs{\mathsf{Q}}_{jk} \,
\bs{q}^i \, \bs{q}^j \, \bs{q}^k \, + \, \bs{\partial}_i
{\bs{\mathsf{J}}^k}_j \, \bs{q}^i \, \bs{q}^j \, \bs{p}_k  \\  
& - \, \frac 12 \, \wt{\bs{\partial}}{}^i\bs{\mathsf{P}}^{jk} \,
\bs{p}_i \, \bs{p}_j \, \bs{p}_k \, - \, \frac 12 \,
\wt{\bs{\partial}}{}^k\bs{\mathsf{Q}}_{ij} \, \bs{q}^i \, \bs{q}^j \,
\bs{p}_k \, - \, \wt{\bs{\partial}}{}^j\bs{\mathsf{J}}^k{}_i \,
\bs{q}^i \, \bs{p}_j \, \bs{p}_k  \Big) \ ,
\end{aligned}\ee
where for simplicity we imposed the $O(d,d)$-invariant boundary condition 
$(\bs{p}_i \, \bs{q}^i)\big|_{T[1]\partial\Sigma_3}=0$.\footnote{This
  boundary condition will be compatible with our further reduction to
  the B-model, but not to the A-model. To be compatible with the
  latter reduction we need to start with a different kinetic term for the large Courant algebroid, in order to obtain the right kinetic term of the Courant sigma-model for the generalized complex structure.}

We reduce the dual coordinates in
\eqref{eq:DoubledDFTSigmaModActionAB} with a gauge fixing
$\wt{\bs{F}}{}^i=0$ and assume that none of the blocks
$\mathsf{P}^{ij}$, ${\mathsf{J}^i}_j$ or $\mathsf{Q}_{ij}$ depend on
$\tx{}_i$. The resulting action is not necessarily an AKSZ action as
it does not satisfy the master equation. Instead we impose the
master equation as a further constraint on the blocks in order to
satisfy the quantization condition, and we define the reduced action
with the constrained blocks, which we write symbolically as 
\be
\mathsf{P}^{ij} \ \xrightarrow{ \ \text{master} \ } \ \pi^{ij} \ ,
\qquad {\mathsf{J}^i}_j \ \xrightarrow{ \ \text{master} \ } \ {J^i}_j
\qquad \mbox{and} \qquad \mathsf{Q}_{ij} \ \xrightarrow{ \
  \text{master} \ } \ \omega_{ij} \ .
\ee
The reduced AKSZ action is given by
\be \begin{aligned}
\bs{\cS}_{\rm{Z}}^{(3)} \, =  \, \int_{T[1]\Sigma_3} \, \dd^3 \hat z \
\Big( & \bs{F}_i \, \bs{D} \bs{X}^i  \, - \, \bs{p}_i \, \bs{D}
\bs{q}^i  \, + \, \bs{\pi}^{ij} \, \bs{F}_i \, \bs{p}_j \, + \,
{\bs{J}^i}_j \, \bs{F}_i \, \bs{q}^j \\
& - \, \frac 12 \, \bs{\partial}_i \bs{\pi}^{jk} \, \bs{q}^i \,
\bs{p}_j \, \bs{p}_k \, - \, \frac 12 \, \bs{\partial}_i
\bs{\omega}_{jk} \, \bs{q}^i \, \bs{q}^j \, \bs{q}^k \, + \,
\bs{\partial}_i {\bs{J}^k}_j \, \bs{q}^i \, \bs{q}^j \, \bs{p}_k \Big) \ .
\end{aligned}\ee
The special property of this AKSZ action is that in the exact gauge
\be
\bs{F}_i \, = \, \mbf D \bs{p}_i \qquad \mbox{and} \qquad \bs{q}^i \, = \, - \mbf D \bs{X}^i 
\ee
it gives the Zucchini action \eqref{eq:ZucchiniAction}:
\be 
\bs{\cS}_{\rm{Z},\rm{gf}}^{(3)} \, = \, \oint_{T[1]\partial\Sigma_3}
\, \dd^2 \hat z \ \Big( \bs{p}_i \, \bs{D}  \bs{X}^i \, + \, \frac 12
\, \bs{\pi}^{ij} \, \bs{p}_i \, \bs{p}_j \, + \, \frac 12 \,
\bs{\omega}_{ij} \, \bs{D} \bs{X}^i \, \bs{D
} \bs{X}^j \, - \, {\bs{J}^i}_j \, \bs{p}_i \, \bs{D} \bs{X}^j \Big) \, = \, \bs{\cS}_{\rm{Z}}^{(2)} \ ,
\ee
on the boundary after the sign flip ${J^i}_j\rightarrow -{J^i}_j$. 

One may naturally expect that the master equation for
$\bs{\cS}_{\rm{Z}}^{(3)}$ will give the constraints of a generalized
complex structure \eqref{eq:GenComplexStrId} as the Zucchini model
does, but this is not precisely true. If $\omega_{ij}$ vanishes then
we get the same identities as those of a generalized complex structure
with $\omega=0$, or if we set $\omega= \pi^{-1}$ then $\dd \omega = 0$
and the term involving $\omega$ vanishes, thus we arrive at the same
AKSZ action. Otherwise the $\omega$ term generally prevents the
constraints from being the identities of a generalized complex structure.

Thus we propose a Courant sigma-model
\be \label{eq:cSpiJ3} \begin{aligned}
\bs{\cS}_{\pi,J}^{(3)} \, =  \, \int_{T[1]\Sigma_3} \, \dd^3 \hat z \
\Big( & \bs{F}_i \, \bs{D} \bs{X}^i  \, - \, \bs{p}_i \, \bs{D}
\bs{q}^i  \, + \, \bs{\pi}^{ij} \, \bs{F}_i \, \bs{p}_j \, + \,
{\bs{J}^i}_j \, \bs{F}_i \, \bs{q}^j \\ & - \, \frac 12 \,
\bs{\partial}_i \bs{\pi}^{jk} \, \bs{q}^i \, \bs{p}_j \, \bs{p}_k \, +
\, \bs{\partial}_i {\bs{J}^k}_j \, \bs{q}^i \, \bs{q}^j \, \bs{p}_k \Big) 
\end{aligned} \ee
for the generalized complex structure
\be \label{eq:GenComplexStrSpecAB}
{\mathbb{J}^I}_J \, = \, \begin{pmatrix} {J^i}_j & \pi^{ij} \\ 0 & -{J^j}_i \end{pmatrix} \ .
\ee
In the language of symplectic dg-geometry this means that the master equation for the Hamiltonian
\be \label{eq:HamABDFTred}
\gamma_{\pi,J} \, = \, \pi^{ij} \, F_i \, p_j \, + \, {J^i}_j \, F_i
\, q^j \, - \, \frac 12 \, \partial_i \pi^{jk} \, q^i \, p_j \, p_k \,
+ \, \partial_i {J^k}_j \, q^i \, q^j \, p_k
\ee
with the symplectic form
\be \label{eq:SymplFormABDFTred}
\omega_{3} \, = \, \dd X^i \w \dd F_i \, + \, \dd q^i \w \dd p_i
\ee
gives the conditions 
\be \begin{aligned} \label{eq:GenComplexStrId2}
\pi^{[i|l} \, \partial_l \pi^{jk]} \, &= \, 0 \ , \\[4pt]
{J^l}_i \, \partial_l \pi^{jk} \, + \, 2 \, \pi^{jl} \, \partial_{[i}
{J^k}_{l]} \, + \, \pi^{kl} \, \partial_l {J^j}_i \, - \, {J^j}_l \, \partial_i \pi^{lk} \, &= \,  0 \ , \\[4pt]
{J^l}_{[i|} \, \partial_l {J^k}_{|j]} \, - \,{J^k}_{l} \, \partial_{[i} {J^{l}}_{j]}  \, &= \, 0 \ .
\end{aligned}\ee
The first identity says that $\pi^{ij}$ satisfies the Poisson condition, the third says ${J^i}_j$ satisfies the integrability condition of the original complex structure, and the second identity is an additional compatibility condition needed to combine them into a generalized complex structure. 

The Hamiltonian $\gamma_{\pi,J}$ defines a Courant algebroid over the
target space $M$ for the generalized complex structures \eqref{eq:GenComplexStrSpecAB} with the Dorfman bracket and anchor
\be
[e_1,e_2]_{\rm{D};\pi,J} \, = \, \{\{ e_1,\gamma_{\pi,J} \},e_2\} \qquad \text{and} \qquad \rho(e) \, = \, \{ e,\{\gamma_{\pi,J}, \, \cdot \, \} \}
\ee
where the functions $e$, $e_1$ and $e_2$ have degree~1. It would be
interesting in its own right to study further this new Courant algebroid structure.

\subsection{Dimensional reductions to the A- and B-models}
\label{sec:DFTtoABmodRed}

The relation of the Courant sigma-model \eqref{eq:cSpiJ3} to the A-model is quite straightforward. If we set $J$ to
zero, and only keep $\pi$ non-zero, the remaining identity from \eqref{eq:GenComplexStrId2} is the
Poisson condition. The resulting AKSZ action is just that of the
contravariant Courant sigma-model, which reduces to the Poisson
sigma-model on its boundary in the exact gauge, and thus to the A-model as well. 

The relation to the B-model is not immediately apparent. Let $\pi$ be zero, and
$J$ non-zero. The remaining identity from \eqref{eq:GenComplexStrId2}
is the integrability condition for the original complex structure
$J$. The Hamiltonian associated to the resulting AKSZ action is
\be \label{eq:HamJ3D}
\gamma_{0,J} \, = \, {J^i}_j \, F_i \, q^j \, + \, \partial_i {J^k}_j
\, q^i \, q^j \, p_k \ ,
\ee
from which a Courant algebroid for a generic complex structure can be derived with the Dorfman bracket and anchor
\be
[e_1,e_2]_{\rm{D};0,J} \, = \, \{\{ e_1,\gamma_{0,J} \},e_2\}   \qquad \text{and} \qquad \rho(e) \, = \, \{ e,\{\gamma_{0,J}, \, \cdot \, \} \}
\ee
respectively, where again $e$, $e_1$ and $e_2$ are degree~1
functions. This structure is similar to that of the Poisson Courant algebroid, which is the derived Courant algebroid for a generic Poisson structure.

We apply the dimensional reduction method introduced in \S\ref{sec:DimRedMeth} on the AKSZ action with the original complex structure solely:
\be \label{eq:AKSZactionDFTJBmod}
\bs{\cS}_{0,J}^{(3)} \, =  \, \int_{T[1]\Sigma_3} \, \dd^3 \hat z \
\big( \bs{F}_i \, \bs{D} \bs{X}^i  \, - \, \bs{p}_i \, \bs{D} \bs{q}^i
\, + \, {\bs{J}^i}_j \, \bs{F}_i \, \bs{q}^j  \, + \, \bs{\partial}_i
{\bs{J}^k}_j \, \bs{q}^i \, \bs{q}^j \, \bs{p}_k \big) \ .
\ee
The reduction method requires that the membrane worldvolume $\Sigma_3$
be a product manifold, hence we apply it in a neighbourhood of the
boundary. For this, choose an open subset $U$ of $\Sigma_3$ which includes $\partial\Sigma_3$:
\be
\Sigma_3\big|_{U} \, = \, \partial\Sigma_3 \times \real^+ \ ,
\ee
where $\real^+$ is the half-line parameterized with coordinate $t$, for
which the $t=0$ points belong to the boundary. Then the worldvolume
$\Sigma_3$ is covered by open sets as
\be \label{eq:Sigma3cover}
\Sigma_3 = U \cup U' \ ,
\ee
where the open set $U'$ does not include the boundary, i.e. it is
contained in the bulk interior
$\Sigma_3\setminus\partial\Sigma_3$. Then the BV symplectic form is
given by the sum of two integrals over the covering sets $U$ and $U'$
as\footnote{We use the same notation for the boundary fields as well
  for brevity, but they are not identical to those used earlier.}
\be\begin{aligned}
\bs{\omega}_{3} \, =  \, \ & \bs{\omega}_{3|U} \, + \, \bs{\omega}_{3|U'} \\[4pt]
 := \, \ & \int_{T[1]U} \, \dd^3 \hat z \ \big(\mbf \delta \bs{X}{}^i \,
 \mbf \delta \bs{F}{}_{i} \, + \, \mbf\delta\bs{q}{}^i \,
 \mbf\delta\bs{p}{}_{i} \big) \, + \,  \int_{T[1]U'} \, \dd^3 \hat z \
 \big(\mbf \delta \bs{X}'{}^i \, \mbf \delta \bs{F}'_{i} \, + \,
 \mbf\delta\bs{q}'{}^{i} \, \mbf\delta\bs{p}{}'_{i} \big) \ ,
\end{aligned}\ee
where we have rescaled the fields $\bs{F}{}_{i}$, $\bs{p}{}_{i}$,
$\bs{F}'_{i}$ and $\bs{p}'_{i}$ with a suitable partition of unity
subordinate to the covering \eqref{eq:Sigma3cover}. These fields are
chosen so that the decomposition of the AKSZ action
\be \begin{aligned} 
\bs{\cS}_{0,J}^{(3)} \, = \, \ & \bs{\cS}_{0,J|U}^{(3)} \, + \, \bs{\cS}_{0,J|U'}^{(3)} \\[4pt]
:= \,  \ & \int_{T[1]U} \, \dd^3 \hat z \ \big( \bs{F}_{i} \, \bs{D}
\bs{X}{}^i  \, - \, \bs{p}_{i} \, \bs{D} \bs{q}{}^i \, + \,
{\bs{J}^i}_j \, \bs{F}_{i} \, \bs{q}{}^j  \, + \, \bs{\partial}_i
{\bs{J}^k}_j \, \bs{q}{}^i \, \bs{q}{}^j \, \bs{p}_{k} \big) \\
 & + \, \int_{T[1]U'} \, \dd^3 \hat z \ \big( \bs{F}'_{i} \, \bs{D}
 \bs{X}'{}^i  \, - \, \bs{p}'_{i} \, \bs{D} \bs{q}'{}^i \, + \,
 {\bs{J}^i}_j \, \bs{F}'_{i} \, \bs{q}'{}^j  \, + \, \bs{\partial}_i
 {\bs{J}^k}_j \, \bs{q}'{}^i \, \bs{q}'{}^j \, \bs{p}'_{k} \big)
\end{aligned}\ee
is independent of the choice of partition of unity.

First we deal with the boundary contributions. They are defined on a
product manifold $\partial\Sigma_3\times\real^+$, so we can apply the
method of \S\ref{sec:DimRedMeth}. We use a uniform notation for an
arbitrary superfield $\mbf\phi$:
\be 
\mbf\phi \, = \, \widehat{\mbf\phi} \, + \, \mbf\phi{}_t \, \theta^t \ ,
\ee
where the component superfields $\widehat{\mbf\phi}$ and
$\mbf\phi{}_t$ do not depend on the odd coordinate $\theta^t$ of $T[1]\real^+$. The integrals over $U$ factorize and we get the BV symplectic form
\be
\bs{\omega}_{3|U;\rm{gf}} \, = \, - \, \oint_{T[1]\partial\Sigma_3} \,
\dd^2 \hat z \ 
\int_{\real^+} \, \dd t \ \big(\mbf \delta
\widehat{\bs{X}}{}^i \, \mbf \delta (\bs{F}{}_{t})_{i} \, + \,
\mbf\delta\widehat{\bs{q}}{}^i \, \mbf\delta(\bs{p}{}_{t})_{i} \big) \ ,
\ee
where we have used a different gauge fixing and also different
antifields than those which were used for the reduction to the A-model: we have set $\bs{X}{}_t^i$ and $\bs{q}{}_t^i$ to zero. The gauge fixed boundary action is
\be\begin{aligned}
\bs{\cS}_{0,J|U;\rm{gf}}^{(3)} \, = \, \oint_{T[1]\partial\Sigma_3} \,
\dd^2 \hat z \
\int_{\real^+} \, \dd t \ \big( & \widehat{\bs{F}}{}_i \, \partial_t
\widehat{\mbf X}{}^i \, + \, \widehat{\bs{p}}{}_i \, \partial_t
\widehat{\mbf q}{}^i \, - \, (\bs{F}_{t})_{i} \, \widehat{\mbf D}
\widehat{\bs{X}}{}^i \, - \, (\bs{p}_{t})_{i} \, \widehat{\mbf D} \widehat{\bs{q}}{}^i \\
 & - \, {\bs{J}^i}_j \, (\bs{F}_{t})_{i} \, \widehat{\bs{q}}{}^j \, + \,
 \bs{\partial}_i {\bs{J}^k}_j \, \widehat{\bs{q}}{}^i \,
 \widehat{\bs{q}}{}^j \, (\bs{p}_{t})_{k} \big) \ .
\end{aligned}\ee
The first two terms determine the boundary conditions. Integrating out
the fields $\widehat{\bs{F}}{}_i$ and $\widehat{\bs{p}}{}_i$ imposes
the condition that the fields $\widehat{\mbf X}{}^i$ and
$\widehat{\mbf q}{}^i$ are independent of $t$, while the zero modes of
$\widehat{\bs{F}}{}_i$ and $\widehat{\bs{p}}{}_i$ on $\real^+$ lead to the condition that $\widehat{\mbf X}{}^i$ and $\widehat{\mbf q}{}^i$ vanish at $t=0$ which means they vanish on the boundary.

We introduce the new notations
\be
\bs{\chi}_i \, = \, - \int_{\real^+} \, \dd t \ (\bs{F}_{t})_{i} \ , \qquad \bs{X}{}^i \, = \, \widehat{\bs{X}}{}^i \ , \qquad \mbf\tx{}_i \, = \, - \int_{\real^+} \, \dd t \ (\bs{p}_{t})_{i} \qquad \text{and} \qquad \ \wt{\bs{\chi}}{}^i \, = \, \widehat{\bs{q}}{}^i \ ,
\ee
and rewrite the BV symplectic form and the boundary AKSZ action with them as
\be
\bs{\omega}_{3|U;\rm{gf}} \, = \,\oint_{T[1]\partial\Sigma_3} \, \dd^2
\hat z \ \big(\mbf \delta \bs{X}{}^i \, \mbf \delta \bs{\chi}{}_i \, +
\, \mbf\delta\mbf\tx{}_i \, \mbf\delta\wt{\bs{\chi}}{}^{i} \big) \ ,
\ee
and
\be \label{eq:BmodNonConstComplStrMinusSign}
\bs{\cS}_{0,J|U;\rm{gf}}^{(3)} \, = \, \oint_{T[1]\partial\Sigma_3} \,
\dd^2 \hat z \ \big(\bs{\chi}_i \, \bs{D}\bs{X}^i \, + \, \mbf
\tx{}_{i} \, \bs{D} \wt{\bs{\chi}}{}^{i} \, + \, {\mbf J^i}_j \,
\bs{\chi}_{i} \, \wt{\bs{\chi}}{}^j \, - \, \bs{\partial}_j
\bs{J}{}^i{}_k \, \mbf \tx{}_i \, \wt{\bs{\chi}}{}^j \, \wt{\bs{\chi}}{}^k \big) \, = \, \bs{\cS}_{J}^{(2)} \ ,
\ee
which give the BV symplectic form corresponding to
\eqref{eq:BmodAKSZ3} and the AKSZ action
\eqref{eq:ComplexStrAKSZaction} for the B-model after the sign flip
$\mbf \tx{}_{i} \rightarrow - \mbf \tx{}_{i}$. Hence the action
$\bs{\cS}_{0,J}^{(3)}$ defined in \eqref{eq:AKSZactionDFTJBmod}
reduces to the B-model action in a neighbourhood of the boundary $\partial\Sigma_3$.

For the bulk contributions, one can gauge fix the bulk fields
independently from the boundary fields using the same fields and
antifields that were used for the reduction to the A-model. The exact
gauge was a gauge fixing on the bulk as well, and not only on the
boundary. Hence if we set $\bs{F}'_i$ and $\bs{q}'{}^i$ to zero, we
get a vanishing bulk action $\bs{\cS}_{0,J|U';\rm{gf}}^{(3)}$, and thus the action $\bs{\cS}_{0,J}^{(3)}$ in \eqref{eq:AKSZactionDFTJBmod} can be reduced entirely to the B-model action on the boundary.

We recall that, in all schemes presented in this paper, the reductions
to the A- and B-models differ significantly: not only are the gauge
choices different, but the antifields are also assigned differently, and the boundary conditions differ as well. But they both appear as boundary AKSZ sigma-models while the bulk fields are gauge fixed completely.

\section{Topological S-duality in generalized complex geometry}
\label{sec:Sduality}

Topological S-duality arises from the physical S-duality of type IIB
superstring theory~\cite{Nekrasov2004b}, and it relates the A- and
B-models on the same Calabi-Yau manifold: D-instanton contributions of
one model correspond to perturbative amplitudes of the other
model. The string couplings $g_{\rm{A}}$ and $g_{\rm{B}}$ of the A- and B-models are related to each other by 
\be \label{eq:gAgB}
g_{\rm{A}} \, = \, \frac{1}{g_{\rm{B}}}
\ee 
and the K{\"a}hler forms $k_{\rm{A}}$ and $k_{\rm{B}}$ of the two theories are also related by 
\be \label{eq:kAkB}
k_{\rm{A}} \, = \, \frac{k_{\rm{B}}}{g_{\rm{B}}} \ .
\ee
In other words, S-duality exchanges the A- and B-models as a
weak/strong coupling duality. As an application of the formalism
developed in this paper, in this section we shall demonstrate how this
duality is
realised geometrically in generalized complex geometry using our Courant algebroids and
AKSZ sigma-models.

\subsection{Duality between Poisson and complex structure Courant algebroids} 

A duality transformation in the language of QP-manifolds is a
transformation of supercoordinates which leaves the symplectic
structure invariant. One of the simplest non-trivial cases is the
renormalization of the fields by a scale transformation: a
coordinate is scaled inversely with respect to its dual coordinate. In the following we study the Courant algebroid for generalized complex structures in this context.

The symplectic form $\omega_{3}$ given by \eqref{eq:SymplFormABDFTred} is left invariant under the scale transformation 
\be \label{eq:ScalingDualS}
p_i \ \longmapsto \ \lambda \, p_i \qquad \text{and} \qquad q^i \ \longmapsto \ \frac{1}{\lambda} \, q^i \ ,
\ee
with a constant parameter $\lambda\in\real$, which transforms the
Hamiltonian $\gamma_{\pi,J}$ from \eqref{eq:HamABDFTred} to
\be \label{eq:HamABLambdaDual}
\gamma_{\pi,J}^{\lambda} \, = \, \lambda \, \pi^{ij} \, F_i \, p_j \,
- \, \frac \lambda2 \, \partial_i \pi^{jk} \, q^i \, p_j \, p_k
\, + \, \frac{1}{\lambda} \, {J^i}_j \, F_i \, q^j  \, + \,
\frac{1}{\lambda} \, \partial_i {J^k}_j \, q^i \, q^j \, p_k \ .
\ee
The scale transformation has no effect on the identities \eqref{eq:GenComplexStrId2}, and $\gamma_{\pi,J}^{\lambda}$ satisfies the master equation as well. 

Now we can take both the large or small $\lambda$ limit. They give different Courant algebroids, namely the Poisson and the complex structure Courant algebroid respectively:
\be
\frac{1}{\lambda} \, \gamma_{0,J} \ \xleftarrow{ \ \lambda \ll 1 \ } \
\gamma_{\pi,J}^{\lambda} \ \xrightarrow{ \ \lambda \gg 1 \ } \ \lambda \, \gamma_{\pi,0} \ ,
\ee
where the Hamiltonian $\gamma_{0,J}$ is defined in \eqref{eq:HamJ3D}
while $\gamma_{\pi,0}$ is defined in \eqref{HamPoisson3D}. After the
limits are taken the parameter $\lambda$ can be scaled back to obtain
the original Hamiltonians which are independent of $\lambda$.

Thus scaling with $\lambda$ introduces a type of weak/strong duality,
which interpolates continuously between Poisson and complex structure Courant
algebroids within the Courant algebroid for generalized complex
geometry, and it exchanges them between the two limits. In the
following we relate this duality to the topological S-duality between
the A- and B-models based on our AKSZ constructions and boundary
reductions from \S\ref{sec:UnifyABDFT}. 

\subsection{Topological S-duality} 

In the following we promote our duality to the level of AKSZ
constructions. We start with the AKSZ action given by the Hamiltonian
$\lambda \, \gamma_{\pi,J}$ defined in \eqref{eq:HamABLambdaDual}:
\be \begin{aligned}
{\bs{\cS}}_{\rm{A/B}}^{(3)} \, =  \, \int_{T[1]\Sigma_3} \, \dd^3 \hat
z \ \Big( & \, \frac{1}{\lambda} \, \bs{F}_i \, \bs{D} \bs{X}^i \, - \,
\frac{1}{\lambda} \, \bs{p}_i \, \bs{D} \bs{q}^i \, + \,
\bs{\pi}^{ij} \, \bs{F}_i \, \bs{p}_j  \, - \, \frac 12 \,
\bs{\partial}_i \bs{\pi}^{jk} \, \bs{q}^i \, \bs{p}_j \, \bs{p}_k \\
& \, + \, {\bs{J}^i}_j \, \bs{F}_i \, \bs{q}^j  \, + \,
\bs{\partial}_i {\bs{J}^k}_j \, \bs{q}^i \, \bs{q}^j \, \bs{p}_k \Big) \ ,
\end{aligned}\ee
where we explicitly introduced an overall constant $1/\lambda$ as a
membrane tension in the definition of the action, which does not
affect the BV quantization of the sigma-model. Now we perform the scaling duality \eqref{eq:ScalingDualS}. Since it leaves the BV symplectic form invariant, the kinetic terms do not change, only the interaction terms. The scale transformed AKSZ action is given by
\be \begin{aligned}
{\bs{\cS}}_{\rm{A/B}}^{\lambda \, (3)} \, =  \, \int_{T[1]\Sigma_3} \,
\dd^3 \hat z \ \Big( & \, \frac{1}{\lambda} \, \bs{F}_i \, \bs{D}
\bs{X}^i \, - \, \frac{1}{\lambda} \, \bs{p}_i \, \bs{D} \bs{q}^i \, +
\, \lambda \, \bs{\pi}^{ij} \, \bs{F}_i \, \bs{p}_j \, - \, 
\frac \lambda2 \, \bs{\partial}_i \bs{\pi}^{jk} \, \bs{q}^i \, \bs{p}_j \, \bs{p}_k \\
& \, + \, \frac{1}{\lambda} \, {\bs{J}^i}_j \, \bs{F}_i \, \bs{q}^j
\, + \, \frac{1}{\lambda} \, \bs{\partial}_i {\bs{J}^k}_j \, \bs{q}^i
\, \bs{q}^j \, \bs{p}_k \Big) \ .
\end{aligned}\ee

The large $\lambda$ limit gives the contravariant Courant sigma-model
without kinetic terms, which reduces to the A-model action given by
\eqref{eq:AmodelAKSZ1997} in the exact gauge in the same way that it reduced in \S\ref{sec:AmodContra}:  
\be
{\bs{\cS}}_{\rm{A/B}}^{\lambda \, (3)} \ \xrightarrow{ \ \lambda \gg 1
\ } \ \frac \lambda2 \, \oint_{T[1]\partial\Sigma_3} \, \dd^2 \hat z \
\bs{\pi}^{ij} \, \bs{p}_i \, \bs{p}_j \ .
\ee
Here $\lambda$ appears as the inverse of the A-model string coupling:
\be \label{eq:LambdaGA}
\lambda \, = \, \frac{1}{g_{\rm{A}}} \ .
\ee

On the other hand, if we take $\lambda$ to be small, we get the AKSZ action of the complex structure Courant algebroid with an overall membrane tension $1/\lambda$, which can be reduced to the B-model on its boundary as in \S\ref{sec:DFTtoABmodRed}:
\be
\bs{\cS}_{\rm{A/B}}^{\lambda \, (3)} \ \xrightarrow{ \ \lambda \ll 1 \
} \ \frac{1}{\lambda} \, \oint_{T[1]\partial\Sigma_3} \, \dd^2 \hat z
\ \Big(\bs{\chi}_i \, \bs{D}\bs{X}^i \, + \, \mbf \tx{}_{i} \, \bs{D}
\wt{\bs{\chi}}{}^{i} \, + \, {\mbf J^i}_j \,
\bs{\chi}_{i} \, \wt{\bs{\chi}}{}^j \, - \, 
\bs{\partial}_j \bs{J}{}^i{}_k \, \mbf \tx{}_i \, \wt{\bs{\chi}}{}^j
\, \wt{\bs{\chi}}{}^k \Big) \ .
\ee
Here $\lambda$ appears as the B-model string coupling this time:
\be
\lambda \, = \, g_{\rm{B}} \ ,
\ee
which together with \eqref{eq:LambdaGA} gives the relation \eqref{eq:gAgB}.
However, it says nothing about the scaling relation of K\"ahler forms
in \eqref{eq:kAkB}. This is due to the fact that the scalings of $\pi$ and $J$ are not fixed to each other by the constraints \eqref{eq:GenComplexStrId2}.   

\section{Conclusions and outlook}
\label{sec:Conc}

In this paper we studied AKSZ formulations of the topological A-
and B-models within the framework of double field theory. We 
presented an observation that the Poisson sigma-model on doubled space
describes both the A- and B-models simultaneously, which are given by
different choices for the doubled Poisson structures. We uplifted
the doubled Poisson sigma-model to the membrane level as a large
contravariant Courant sigma-model, which is the three-dimensional AKSZ
sigma-model description of a doubled Poisson sigma-model. We 
showed that it reduces to the doubled Poisson sigma-model on the
boundary if we use the exact gauge, and we have expounded some
particular cases given by the reduction which are relevant in AKSZ theory.

We applied the projection to DFT vectors on the large
contravariant Courant sigma-model of double field theory, which halves
the number of ghost number~1 fields, and we reduced the dual
coordinates, which led to the introduction of the Courant sigma-model
of a particular class of generalized complex structures, and also its
corresponding Courant algebroid. We pointed out that it gives the
Zucchini model on the boundary in the exact gauge. We also studied two
marginal cases, the purely Poisson and purely complex structure cases,
which were reduced to the A- and B-models on their boundaries respectively. 

We also proposed an S-duality between Poisson and complex structure
Courant algebroids, which originates back to the generalized complex
structure, in which they lie as different marginal limits. We promoted
the duality to the AKSZ level, and related it to topological S-duality
of the A- and B-models.

It would be interesting to study further the appearance of generalized
geometry and double field theory in the context of the A- and B-models
as they are defined originally in standard (not generalized)
Calabi-Yau manifolds. The double field theory formulation of the A-
and B-models also allows for the introduction of both geometric and
non-geometric fluxes, which would be a further development in the
study of its physical relevance, particularly in the context of
topological string theory. The fluxes correspond to twist deformations
of the proposed Courant algebroids which lead to the introduction of
twists of the generalized and original complex structures, which is
another avenue for further investigation. Another direction would be
to find a Courant algebroid which gives the identities of a general
version of generalized complex structure, and to relate it to the
double field theory formulations of the A- and B-models. Our S-duality
gives a continuous mapping between the A- and B-models, so it would be
interesting to investigate whether the intermediate membrane theory
has a clearer physical relevance. A surprising observation is that our
S-duality arises from the T-duality inspired generalized complex geometry,
thus it raises the question as to whether there is a physical origin
behind this relation or whether it is just a coincidence found in the
topological field theories.

\section*{Acknowledgments}

This work was completed while R.J.S. was visiting the Institut
Montpelli\'erain Alexander Grothen{-}dieck in Montpellier, France during
May/June 2018; he warmly thanks the staff there for hospitality and
for providing a stimulating environment, and in particular Damien
Calaque for the invitation and numerous discussions.
This work was supported by the Action MP1405 QSPACE, funded by the European Cooperation in Science and Technology (COST).
The work of Z.K. was supported by the Hungarian Research Fund (OTKA). 
The work of R.J.S. was supported by the Consolidated Grant
ST/P000363/1 from the UK Science and Technology Facilities Council, and
by the
Institut
Universitaire de France.

\bigskip

\appendix

\section{Differential calculus of graded functionals}
\label{sec:appGradedFormulas}

A nice and elaborate summary of formulas from differential calculus on graded manifolds can be found in~\cite{Ikeda2012}. In the following we rely on this treatment and only review formulas with regard to graded functionals. 

The space of superfields is once again the mapping space 
\be
\mbf\cM = {\sf Map}\big(T[1]\Sigma_d\,,\,\cM\big) \ ,
\ee
and the superfield coordinates on $\mbf\cM$ are defined via the coordinates $\hat z^{\hat\mu}\in T[1]\Sigma_d$, $\hat X^{\hat\imath}\in \cM$ as 
\be
\hat{\mbf X}{}^{\hat\imath}(\hat z) \, = \, \mbf\phi^*(\hat
X^{\hat\imath}\,)(\hat z) \  ,
\ee
where $\mbf\phi\in\mbf\cM$ is an arbitrary superfield. We use the
notation $|\hat\imath|$ for the degree of $\hat X^{\hat\imath}$, and
also for the ghost number of $\hat{\mbf X}{}^{\hat\imath}$. 

A graded functional\footnote{We only consider non-local graded
  functionals, hence the kernel function $F( \hat{\mbf X}\big)$ can be
  taken to be an ordinary graded function of $\hat{\mbf X}$.} of the
superfields $\hat{\mbf X}{}^{\hat\imath}(\hat z)$ is defined by a
function $F( \hat{X})$ on $\cM$ as
\be
{\mbf F} \, = \, \int_{T[1]\Sigma_{d}} \, \dd^d \hat z \ \mathrm{ev}^*
(F) \, = \, \int_{T[1]\Sigma_{d}} \, \dd^d \hat z \ F\big( \hat{\mbf X}(\hat z) \big) \ ,
\ee 
and it has ghost number $|F|-d$, where $|F|$ denotes the ghost number
of $F\big( \hat{\mbf X}(\hat z) \big)$. The definition of a graded
functional  $n$-form $\bs{\alpha}$ with ghost number $|\alpha|-d$ is
analogous and given by an $n$-form $\alpha$ on $\cM$ as
\be \begin{aligned}
\bs{\alpha}\, =& \, \int_{T[1]\Sigma_{d}} \, \dd^d \hat z \
\mathrm{ev}^* (\alpha) \\[4pt] \, =& \,  \int_{T[1]\Sigma_{d}} \,
\dd^d \hat z \ \mbf\delta \hat{\bs{X}}{}^{\hat\imath_1}(\hat{z})
\cdots \mbf\delta \hat{\bs{X}}{}^{\hat\imath_n}(\hat{z}) \,
\alpha_{\hat\imath_1\cdots\hat\imath_n}\big(\hat{\bs{X}}(\hat{z})\big)
\\[4pt] \, :=& \, \int_{T[1]\Sigma_{d}} \, \dd^d \hat z \ \alpha\big( \hat{\mbf X}(\hat z) \big) \ .
\end{aligned}\ee
The exterior product of two graded functional forms $\mbf\alpha$ and
$\mbf\beta$ also gives a graded functional form with ghost number
$|\alpha|+|\beta|-d$ which reads as
\be
\mbf\alpha \mbf\w \mbf\beta \, = \, \int_{T[1]\Sigma_{d}} \, \dd^d
\hat z \ \mathrm{ev}^* (\alpha \w \beta) \, = \, \int_{T[1]\Sigma_{d}}
\, \dd^d \hat z \ \alpha\big( \hat{\mbf X}(\hat z) \big) \, \beta\big( \hat{\mbf X}(\hat z) \big) \ .
 \ee
As can be seen, it depends on the integration measure. The one-form
local basis elements have ghost number $|\hat \imath|+1-d$ and they
are given by
\be
\mbf\delta \hat{\mbf X}{}^{\hat\imath} \, = \, \int_{T[1]\Sigma_{d}}
\, \dd^d \hat z \ \mathrm{ev}^* (\dd \hat X {}^{\hat\imath}) \, = \,
\int_{T[1]\Sigma_{d}} \, \dd^d \hat z \ \mbf\delta \hat{\mbf X}{}^{\hat\imath}(\hat z) \ .
\ee
We can write the $n$-form $\bs{\alpha}$ with the exterior product as
\be
\bs{\alpha}\, = \,  \mbf\delta \hat{\bs{X}}{}^{\hat\imath_1} \mbf\w
\cdots \mbf\w \mbf\delta \hat{\bs{X}}{}^{\hat\imath_n} \mbf\w
\mbf\alpha{}_{\hat\imath_1\cdots\hat\imath_n} \ ,
\ee
where the scalar functional is given by
\be
\mbf\alpha{}_{\hat\imath_1\cdots\hat\imath_n} \, = \,
\int_{T[1]\Sigma_{d}} \, \dd^d \hat z \ \alpha_{\hat\imath_1\cdots\hat\imath_n}\big(\hat{\bs{X}}(\hat z)\big) \ .
\ee
 
A graded vector functional $\mbf V$ with ghost number $|V|-d$ can be written in the form
\be
\mbf V \, = \, \int_{T[1]\Sigma_{d}} \, \dd^d \hat z \ V^{\hat\imath}\big( \hat{\mbf X}(\hat z) \big) \, \frac{\mbf\delta}{\mbf\delta \hat{\mbf X}{}^{\hat\imath}(\hat z)} \ ,
\ee 
which acts as a left functional derivative
\be
\bs{\overset{\shortrightarrow}{V}} \bs{F} \, = \,
\int_{T[1]\Sigma_{d}} \, \dd^d \hat z \ V^{\hat\imath}\big( \hat{\mbf X}(\hat z) \big) \, \frac{\rdelta \bs{F}}{\delta \hat{\mbf X}{}^{\hat\imath}(\hat z)} 
\ee 
on graded functionals $\bs{F}$, and is defined as
\be
\lim_{\epsilon\rightarrow 0} \, \frac{\bs{F}[\hat{\bs{X}} + \epsilon
  \, \hat{\bs{\eta}}]\, - \, \bs{F}[\hat{\bs{X}}]}{\epsilon} \, =: \,
\int_{T[1]\Sigma_{d}} \, \dd^d \hat z \ \hat{\bs{\eta}}{}^{\hat\imath}(\hat z) \, \frac{\rdelta \bs{F}}{\delta \hat{\mbf X}{}^{\hat\imath}(\hat z)} \ ,
\ee
where $\hat{\bs{\eta}}{}^{\hat\imath}$ is an arbitrary superfield with
the same degree as $\hat{\bs{X}}{}^{\hat\imath}$. The functional
derivative with respect to $\hat{\mbf X}{}^{\hat\imath}(\hat z)$ has
ghost number $-|\hat\imath|+d$. As graded functionals are non-local,
the functional derivatives are given by ordinary derivatives of the
kernel function as
\be
\frac{\rdelta \bs{F}}{\delta \hat{\mbf X}{}^{\hat\imath}(\hat z)} \, =
\, \frac{\rd {F}}{\partial \hat{X}{}^{\hat\imath}} \bigg|_{\hat{\mbf X}{}^{\hat\imath}(\hat z)} \ .
\ee

The interior product is given by contraction with the graded vector functional $\mbf V$:
\be
\iota_{\mbf V} \, = \, \int_{T[1]\Sigma_{d}} \, \dd^d \hat z \ V^{\hat\imath}\big( \hat{\mbf X}(\hat z) \big) \, \frac{\rdelta}{\delta\, ( \mbf\delta\hat{\mbf X}{}^{\hat\imath}(\hat z))} \ ,
\ee
which acts on exterior products as
\be
\iota_{\mbf V}  (\mbf\alpha \mbf\w \mbf\beta) \, = \, \iota_{\mbf V}
\mbf\alpha \mbf\w \mbf\beta \, + \, (-1)^{(|V|+1)\,|\alpha|} \, \mbf\alpha \mbf\w \iota_{\mbf V}\mbf\beta \ .
\ee

The de Rham differential can be written in the form
\be
\mbf\delta \, = \, \int_{T[1]\Sigma_{d}} \, \dd^d \hat z \ \mbf\delta\hat{\mbf X}{}^{\hat\imath}(\hat z) \, \frac{\rdelta }{\delta \hat{\mbf X}{}^{\hat\imath}(\hat z)} \ .
\ee 
It has the following properties:
\be \begin{aligned}
\mbf\delta \mbf F \, &= \,  \lim_{\epsilon\rightarrow 0} \,
\frac{\bs{F}[\hat{\bs{X}} + \epsilon \, \mbf\delta\hat{\bs{X}}]\, - \,
  \bs{F}[\hat{\bs{X}}]}{\epsilon} \ , \\[4pt]
\mbf\delta^2 \, &=  \, 0 \ , \\[4pt]
\mbf\delta (\mbf\alpha \mbf\w \mbf\beta) \, &= \, \mbf\delta
\mbf\alpha \mbf\w \mbf\beta \, + \, (-1)^{|\alpha|} \, \mbf\alpha
\mbf\w \mbf\delta \mbf\beta \ , \\[4pt]
\mbf\delta \mbf\alpha \, &=  \, \int_{T[1]\Sigma_{d}} \, \dd^d \hat z \ \mathrm{ev}^* (\dd\alpha) \ .
\end{aligned} \ee

The symplectic structure on the target dg-manifold $\cM$ has degree $d+1$ and it can be written in the form
\be
\omega=(-1)^{(d+1)\,|a|} \, \dd q^a \w \dd p_a \ ,
\ee
where $q^a$ and $p_a$ are local Darboux coordinates such that $|q^a|+|p_a|=d-1$. The BV symplectic form $\mbf\omega$ with ghost number $1$ is defined by    
\be
\mbf\omega \, = \, \int_{T[1]\Sigma_{d}} \, \dd^d \hat z \
\mathrm{ev}^* (\omega) \, = \, \int_{T[1]\Sigma_{d}} \, \dd^d \hat z \
(-1)^{(d+1)\,|q^a|} \, \mbf\delta \bs{q}^a(\hat z) \mbf\w \mbf\delta
\bs{p}_a(\hat z) \ .
\ee
The definition of the Hamiltonian vector field of a graded functional
$\mbf F$ is given by the expression
\be
\iota_{\mbf X{}_{\mbf F}} \mbf\omega \, := \, \mbf\delta \mbf F \ ,
\ee 
and it has ghost number $|F|-d+1$. The solution to this equation 
\be
\mbf X{}_{\mbf F} \, = \, (-1)^{|F|+d} \, \int_{T[1]\Sigma_{d}} \,
\dd^d \hat z \ \bigg( \frac{\bs{F} \ldelta }{\delta \bs{q}^a(\hat z)}
\, \frac{ \mbf\delta }{\mbf\delta \bs{p}_a(\hat z)} \, - \,
(-1)^{|q^a|\,|p_a|} \, \frac{\bs{F} \ldelta }{\delta \bs{p}_a(\hat z)}
\, \frac{ \mbf\delta }{\mbf\delta \bs{q}^a(\hat z)} \bigg) 
\ee
is used to define the BV bracket of graded functionals $\mbf F$ and
$\mbf G$ as
\be \begin{aligned}
\bv{\mbf F}{\mbf G} \, &= \, (-1)^{|F|+d} \, \bs{
  \overset{\shortrightarrow}{X}}{}_{\mbf F} \mbf G \\[4pt] \, &= \,
(-1)^{|F|+d} \, \iota_{\mbf X{}_{\mbf F}} \mbf\delta \mbf G \\[4pt] \, &= \,
 (-1)^{|F|+d+1} \, \iota_{\mbf X{}_{\mbf F}} \iota_{\mbf X{}_{\mbf G}} \mbf\omega \\[4pt]
\, &= \, \int_{T[1]\Sigma_{d}} \, \dd^d \hat z \ \bigg( \frac{\bs{F}
  \ldelta }{\delta \bs{q}^a(\hat z)} \, \frac{ \rdelta \mbf G }{\delta
  \bs{p}_a(\hat z)} \, - \, (-1)^{|q^a|\,|p_a|} \, \frac{\bs{F}
  \ldelta }{\delta \bs{p}_a(\hat z)} \, \frac{ \rdelta \mbf G }{\delta \bs{q}^a(\hat z)} \bigg) \ .
\end{aligned}\ee
The BV bracket has the following properties:
\be \begin{aligned}
\bv{\mbf F}{\mbf G} \, &= \, - (-1)^{(|F|+d+1)\,(|G|+d+1)} \, \bv{\mbf G}{\mbf F} \ , \\[4pt]
\bv{\mbf F}{\mbf G \, \mbf H} \, &= \, \bv{\mbf F}{\mbf G} \, \mbf H
\, + \, (-1)^{(|F|+d+1)\,|G|} \, \mbf G \, \bv{\mbf F}{\mbf H} \ , \\[4pt]
\bv{\mbf F}{\bv{\mbf G}{\mbf H}} \, &= \, \bv{\bv{\mbf F}{\mbf G}}{\mbf H} \, + \, (-1)^{(|F|+d+1)\,(|G|+d+1)} \, \bv{\mbf G}{\bv{\mbf F}{\mbf H}} \ .
\end{aligned} \ee


\bigskip

\begin{thebibliography}{99}

\bibitem{Witten1988a}
E.~Witten, ``Topological quantum field theory,'' Commun.~Math.~Phys. {\bf 117} (1988) 353--386. 

\bibitem{Witten1988b}
E.~Witten, ``Topological sigma-models,'' Commun.~Math.~Phys. {\bf 118} (1988) 411--449.

\bibitem{Witten1991}
E.~Witten, ``Mirror manifolds and topological field theory,'' AMS/IP Stud.\ Adv.\ Math.\  {\bf 9} (1998) 121--160
[arXiv:hep-th/9112056].

\bibitem{Bershadsky1994}
M.~Bershadsky, S.~Cecotti, H.~Ooguri and C.~Vafa, ``Kodaira-Spencer theory of gravity and exact results for quantum string amplitudes,'' Commun. Math. Phys.
{\bf 165} (1994) 311--428
[arXiv:hep-th/9309140].

\bibitem{AKSZ1997}	
M.~Alexandrov, M.~Kontsevich, A.~Schwarz and O.~Zaboronsky,
``The geometry of the master equation and topological quantum field theory,'' Int.~J.~Mod.~Phys.~A {\bf 12} (1997) 1405--1429
[arXiv:hep-th/9502010].

\bibitem{Cattaneo2001}
  A. S. Cattaneo and G. Felder,
  ``On the AKSZ formulation of the Poisson sigma-model,'' Lett.~Math.~Phys.~{\bf 56} (2001) 163--179
	[arXiv:math.QA/0102108].
	
\bibitem{Bouwknegt:2011vn}
  P.~Bouwknegt and B.~Jur\v{c}o,
  ``AKSZ construction of topological open $p$-brane action and Nambu brackets,''
  Rev.\ Math.\ Phys.\  {\bf 25} (2013) 1330004
  [arXiv:1110.0134 [math-ph]].
	
\bibitem{Ikeda2012}
N.~Ikeda,
	``Lectures on AKSZ sigma-models for physicists,''
	in: {\em Noncommutative Geometry and Physics 4},
        eds. Y.~Maeda, H.~Moriyoshi, M.~Kotani and S.~Watamura (World
        Scientific, 2017) 79--170
	[arXiv:1204.3714 [hep-th]].
	
\bibitem{Hull2005}
  C.~M.~Hull,
  ``A geometry for non-geometric string backgrounds,''
  JHEP {\bf 0510} (2005) 065
[arXiv:hep-th/0406102].

\bibitem{Shelton2005}
  J.~Shelton, W.~Taylor and B.~Wecht,
  ``Non-geometric flux compactifications,''
  JHEP {\bf 0510} (2005) 085
	[arXiv:hep-th/0508133].

\bibitem{Blumenhagen2012}
  R.~Blumenhagen, A.~Deser, E.~Plauschinn and F.~Rennecke,
  ``Bianchi identities for non-geometric fluxes: From quasi-Poisson structures to Courant algebroids,''
  Fortsch. Phys. {\bf 60} (2012) 1217--1228
	[arXiv:1205.1522 [hep-th]].
	
\bibitem{Heller2016}	
M.~A.~Heller, N.~Ikeda and S.~Watamura, ``Unified picture of non-geometric fluxes and T-duality
in double field theory via graded symplectic manifolds,'' JHEP~{\bf 1702} (2017) 078
[arXiv:1611.08346 [hep-th]].

\bibitem{Roytenberg2002}
  D.~Roytenberg,
  ``Courant algebroids, derived brackets and even symplectic supermanifolds,''
	PhD Thesis, University of California at Berkeley [arXiv:math.DG/9910078].
	
	\bibitem{Park2000}
  J.-S.~Park,
  ``Topological open $p$-branes,''
  in: {\em Symplectic Geometry and Mirror Symmetry}, eds. K.~Fukaya, Y.-G.~Oh, K.~Ono and G.~Tian (World Scientific, 2001) 311--384
  [arXiv:hep-th/0012141].
	
\bibitem{Ikeda2003}
 N.~Ikeda, ``Chern-Simons gauge theory coupled with BF-theory,'' Int.~J.~Mod.~Phys.~A {\bf 18} (2003) 2689--2702
[arXiv:hep-th/0203043].

\bibitem{Roytenberg2002b}
 D.~Roytenberg,
  ``On the structure of graded symplectic supermanifolds and Courant algebroids,''
   Contemp. Math. {\bf 315} (2002) 169--186 [arXiv:math.SG/0203110].
	
\bibitem{Hofman2002}	
C.~Hofman and J.-S.~Park, ``Topological open membranes,'' arXiv:hep-th/0209148.	
	
\bibitem{Hofman2002a}
  C.~Hofman and J.-S.~Park,
  ``BV quantization of topological open membranes,''
  Commun.\ Math.\ Phys.\  {\bf 249} (2004) 249--271
  [arXiv:hep-th/0209214].
	
\bibitem{Roytenberg2007}
  D.~Roytenberg,
  ``AKSZ--BV formalism and Courant algebroid-induced topological field theories,'' Lett.~Math.~Phys.~{\bf 79} (2007) 143--159
	[arXiv:hep-th/0608150].
	
	\bibitem{Halmagyi}
	N.~Halmagyi,
	``Non-geometric backgrounds and the first order string sigma-model,''
	arXiv:0906.2891 [hep-th].
	
\bibitem{Richard2012}
D.~Mylonas, P.~Schupp and R.~J.~Szabo,
``Membrane sigma-models and quantization of non-geometric flux backgrounds,'' JHEP {\bf 1209} (2012) 012
	[arXiv:1207.0926 [hep-th]].
	
\bibitem{Chatzistavrakidis2015}
  A.~Chatzistavrakidis, L.~Jonke and O.~Lechtenfeld,
  ``Sigma-models for genuinely non-geometric backgrounds,''
  JHEP {\bf 1511} (2015) 182
  [arXiv:1505.05457 [hep-th]].

\bibitem{Szabo2018}
  R.~J.~Szabo,
  ``Higher quantum geometry and non-geometric string theory,''
  arXiv:1803.08861 [hep-th].

	\bibitem{doubled3}
	W.~Siegel,
	``Two vierbein formalism for string inspired axionic gravity,''
	Phys.\ Rev.\ D {\bf 47} (1993) 5453--5459
	[arXiv:hep-th/9302036].
	
	\bibitem{Siegel:1993th}
	W.~Siegel,
	``Superspace duality in low-energy superstrings,''
	Phys.\ Rev.\ D {\bf 48} (1993) 2826--2837
	[arXiv:hep-th/9305073].

	\bibitem{dft1}
	C.~M.~Hull and B.~Zwiebach,
	``Double field theory,''
	JHEP {\bf 0909} (2009) 099
	[arXiv:0904.4664 [hep-th]].
	
\bibitem{Aldazabal2013}
   G.~Aldazabal, D.~Marqu\'es and C.~N\'u\~nez,
  ``Double field theory: A pedagogical review,''
  Class.~Quant.~Grav.~{\bf 30} (2013) 163001
	[arXiv:1305.1907 [hep-th]].
	
\bibitem{Berman2014}
   D.~S.~Berman and D.~C.~Thompson,
  ``Duality symmetric string and M-theory,''  Phys.~Rept.~{\bf 566} (2014) 1--60
	[arXiv:1306.2643 [hep-th]].

\bibitem{Hohm2013}
    O.~Hohm, D.~L\"ust and B.~Zwiebach,
  ``The spacetime of double field theory: Review, remarks, and outlook,'' Fortsch.~Phys.~{\bf 61} (2013) 926--966
	[arXiv:1309.2977 [hep-th]].
	
	\bibitem{Hull:2009zb}
	C.~M.~Hull and B.~Zwiebach,
	``The gauge algebra of double field theory and Courant brackets,''
	JHEP {\bf 0909} (2009) 090
	[arXiv:0908.1792 [hep-th]].
	
	\bibitem{Vaisman:2012ke}
	I.~Vaisman,
	``On the geometry of double field theory,''
	J.\ Math.\ Phys.\  {\bf 53} (2012) 033509
	[arXiv:1203.0836 [math.DG]].
	
\bibitem{Deser2015}
 A.~Deser and J.~Stasheff, ``Even symplectic supermanifolds and double field theory,'' Commun.~Math.~Phys.~{\bf 339} (2015) 1003--1020
 [arXiv:1406.3601 [math-ph]].

\bibitem{Deser2016}
 A.~Deser and C.~Saemann, ``Extended Riemannian geometry I: Local double field theory,'' 
arXiv:1611.02772 [hep-th].

	\bibitem{Freidel:2017yuv}
	L.~Freidel, F.~J.~Rudolph and D.~Svoboda,
	``Generalised kinematics for double field theory,''
	JHEP {\bf 1711} (2017) 175
	[arXiv:1706.07089 [hep-th]].

\bibitem{Richard2018}
A.~Chatzistavrakidis, L.~Jonke, F.~S.~Khoo and R.~J.~Szabo,
``Double field theory and membrane sigma-models,''
	arXiv:1802.07003 [hep-th].
	
\bibitem{Svoboda:2018rci}
D.~Svoboda,
``Algebroid structures on para-Hermitian manifolds,''
arXiv:1802.08180 [math.DG].

\bibitem{Kapustin:2003sg}
  A.~Kapustin,
  ``Topological strings on noncommutative manifolds,''
  Int.\ J.\ Geom.\ Meth.\ Mod.\ Phys.\  {\bf 1} (2004) 49--81
  [arXiv:hep-th/0310057].

\bibitem{Kapustin:2004gv}
  A.~Kapustin and Y.~Li,
  ``Topological sigma-models with $H$-flux and twisted generalized complex manifolds,''
  Adv.\ Theor.\ Math.\ Phys.\  {\bf 11} (2007)  269--290
  [arXiv:hep-th/0407249].

\bibitem{Pestun:2005rp}
  V.~Pestun and E.~Witten,
  ``The Hitchin functionals and the topological B-model at one loop,''
  Lett.\ Math.\ Phys.\  {\bf 74} (2005) 21--51
  [arXiv:hep-th/0503083].
	
\bibitem{Zucchini2004}
R.~Zucchini, ``A sigma-model field theoretic realization of Hitchin's generalized complex geometry,'' JHEP {\bf 0411} (2004) 045
[arXiv:hep-th/0409181].

\bibitem{Zucchini2005}
R.~Zucchini, ``Generalized complex geometry, generalized branes and the Hitchin sigma-model,'' JHEP {\bf 0503} (2005) 022
[arXiv:hep-th/0501062].
	
 \bibitem{Pestun2006}
   V.~Pestun,
   ``Topological strings in generalized complex space,'' 
   Adv.\ Theor.\ Math.\ Phys.\  {\bf 11} (2007) 399--450
 [arXiv:hep-th/0603145].

\bibitem{Ikeda2007}
N.~Ikeda and T.~Tokunaga, ``An alternative topological field theory of generalized complex geometry,'' JHEP {\bf 0709} (2007) 009
[arXiv:0704.1015 [hep-th]].

\bibitem{Stojevic2005}
  V.~Stojevic,
  ``Topological A-type models with flux,''
  JHEP {\bf 0805} (2008) 023 
	[arXiv:0801.1160 [hep-th]].
	
 \bibitem{Bessho2015}
   T.~Bessho, M.~A.~Heller, N.~Ikeda and S.~Watamura,
   ``Topological membranes, current algebras and $H$-flux--$R$-flux duality based on Courant algebroids,'' JHEP {\bf 1604} (2016) 170
 [arXiv:1511.03425 [hep-th]].

\bibitem{Nekrasov2004b}
N.~Nekrasov, H.~Ooguri and C.~Vafa, ``S-duality and topological strings,'' JHEP {\bf 0410} (2004) 009
[arXiv:hep-th/0403167]

 \bibitem{Cattaneo2009}
   A.~S.~Cattaneo, J.~Qiu and M.~Zabzine,
   ``$2D$ and $3D$ topological field theories for generalized complex geometry,'' Adv.~Theor.~Math.~Phys.~{\bf 14} (2010) 695--725
 [arXiv:0911.0993 [hep-th]].

\bibitem{Kokenyesi2018}
Z.~K\"ok\'enyesi, A.~Sinkovics and R.~J.~Szabo, ``AKSZ constructions for topological membranes on $G_2$-manifolds,''
Fortsch. Phys. {\bf 66} (2018) 1800018
	[arXiv:1802.04581 [hep-th]].

\bibitem{Batalin1981}
  I.~A.~Batalin and G.~A.~Vilkovisky,
  ``Gauge algebra and quantization,''
  Phys.\ Lett.\  B {\bf 102} (1981) 27--31.

\bibitem{Mnev2008}
  P.~Mnev,
  ``Discrete BF-theory,''
  arXiv:0809.1160 [hep-th].
	
\bibitem{Ikeda1993}
  N.~Ikeda,
  ``Two-dimensional gravity and nonlinear gauge theory,''
  Ann. Phys.\  {\bf 235} (1994) 435--464
  [arXiv:hep-th/9312059].
  
\bibitem{Schaller1994}
  P.~Schaller and T.~Strobl,
  ``Poisson structure induced (topological) field theories,''
  Mod.\ Phys.\ Lett.\ A {\bf 9} (1994) 3129--3136
  [arXiv:hep-th/9405110].

\bibitem{CattFeld1999}
A.~S.~Cattaneo and G.~Felder, 
``A path integral approach to the Kontsevich quantization formula,''
Commun.~Math.~Phys.~{\bf 212} (2000) 591--611
[arXiv:math.QA/9902090].
		
 \bibitem{Hofman2002b}
   C.~Hofman,
   ``On the open-closed B-model,'' JHEP {\bf 0311} (2003) 069
 [arXiv:hep-th/0204157].

\bibitem{Hitchin2004}
  N.~Hitchin,
  ``Generalized Calabi-Yau manifolds,''
  Quart.\ J.\ Math.\  {\bf 54} (2003) 281--308
  [arXiv:math.DG/0209099].
  
\bibitem{Gualtieri2003}
  M.~Gualtieri,
  ``Generalized complex geometry,''
  Ann. Math. {\bf 174} (2011) 75--123 
  [math.DG/0401221].

\bibitem{Asakawa2014}
  T.~Asakawa, H.~Muraki, S.~Sasa and S.~Watamura,
  ``Poisson-generalized geometry and $R$-flux,''
  Int.\ J.\ Mod.\ Phys.\ A {\bf 30} (2015) 1550097
  [arXiv:1408.2649 [hep-th]].

	\bibitem{Blumenhagen2010}
	R.~Blumenhagen and E.~Plauschinn,
	``Nonassociative gravity in string theory?,''
	J.\ Phys.\ A {\bf 44} (2011) 015401
	[arXiv:1010.1263 [hep-th]].

\bibitem{Lust2010}
D.~L\"ust,
``T-duality and closed string noncommutative (doubled) geometry,''
JHEP {\bf 1012} (2010) 084
[arXiv:1010.1361 [hep-th]].

	\bibitem{Blumenhagen2011}
	R.~Blumenhagen, A.~Deser, D.~L\"ust, E.~Plauschinn and F.~Rennecke,
	``Non-geometric fluxes, asymmetric strings and nonassociative geometry,''
	J.\ Phys.\ A {\bf 44} (2011) 385401
	[arXiv:1106.0316 [hep-th]].
	
  
\end{thebibliography}
\end{document}